\documentclass[fleqn,usenatbib]{mnras}
\usepackage{graphicx}
\usepackage{threeparttable}
\usepackage{multirow}

\usepackage{amsmath}
\usepackage{amssymb}

\newcommand\dalpha{\Delta\alpha/\alpha}
\newcommand\oiii{[{O$\;${\small\rmfamily \uppercase\expandafter{\romannumeral 3}}\relax}]\ }

\title{Time Variation of Fine-Structure Constant Constrained by [O III] Emission-Lines at $1.1<z<3.7$}

\author[Ge Li et al.]{
Ge Li$^{1}$\thanks{E-mail: lige413@mail.ustc.edu.cn}, Luming Sun$^{2}$, Xiangjun Chen$^{1}$, and Hongyan Zhou$^{3,1}$
\\
$^{1}$Department of Modern Physics,University of Science and Technology of China, Hefei, Anhui, 230026, China.\\
$^{2}$Department of Physics, Anhui Normal University, Jiuhuanan Road 189, Wuhu 241002, China\\
$^{3}$Polar Research Institute of China, 451 Jinqiao Road, Shanghai, China
}

\date{Accepted XXX. Received YYY; in original form ZZZ}

\pubyear{}

\begin{document}
\label{firstpage}
\pagerange{\pageref{firstpage}--\pageref{lastpage}}
\maketitle

\begin{abstract}
  [O III]$\lambda\lambda$4960,5008 doublet are often the strongest narrow emission lines in starburst galaxies and quasi-stellar objects (QSOs), and thus are a promising probe to possible variation of the fine-structure constant $\alpha$ over cosmic time.
  Previous such studies using QSOs optical spectra were limited to $z<1$.
  In this work, we constructed a sample of 40 spectra of Ly$\alpha$ emitting galaxies (LAEs) and a sample of 46 spectra of QSOs at $1.09<z<3.73$ using the VLT/X-Shooter near-infrared spectra publicly available.
  We measured the wavelength ratios of the two components of the spin-orbit doublet and accordingly calculated $\alpha(z)$ using two methods.
  Analysis on all of the 86 spectra yielded $\dalpha=(-3\pm6)\times10^{-5}$ with respect to the laboratory $\alpha$ measurements, consistent with no variation over the explored time interval.
  If assuming a uniform variation rate, we obtained $\alpha^{-1}{\rm d}\alpha/{\rm d}t = (-3\pm6)\times10^{-15}$ yr$^{-1}$ within the last 12 Gyrs.
  Extensive tests indicate that $\alpha$ variation could be better constrained using starburst galaxies' spectra than using QSO spectra in future studies.
\end{abstract}

\section{Introduction}

\cite{Dirac1937} proposed that fundamental physical constants may change with the evolution of the universe.
The fine structure constant $\alpha$ is defined as:
\begin{equation}
\rm \alpha= \frac{e^2}{4\pi\varepsilon_0 \hbar c},
\end{equation}
where $e$ is the electron charge, $\varepsilon_0$ is the permittivity of free space, $\hbar$ is the Planck constant, and $c$ is the speed of light.
It is a fundamental physical constant charactering the strength of the electromagnetic interaction between elementary charged particles.
It also quantifies the gap in the fine structure of the spectral lines of atom.
This gap is proportional to the energy of main level by a factor of $\alpha^2$ \citep[e.g.,][]{Dzuba1999,Uzan2003}.
Thus variation of $\alpha$ over time could be directly constrained by comparing the wavelengths of fine-structure splitting of atomic lines from another epoch with that from today.
Possible variation of $\alpha$ over time has been studied with the help of measurements from laboratories, geology and astronomy \citep[see review by][]{Uzan2003,Uzan2011}.
Among these, the studies involving astronomical spectra can provide possible variation of $\alpha$ with the longest lookback time and the greatest spatial spans, in which case trivial variation of $\alpha$ with spacetime might accumulate to a measurable level.

Observational studies using astronomical spectroscopy began in the 1950s \citep[e.g.,][]{Savedoff1956}.
From the 1960s, doublet lines of \oiii $\lambda\lambda$4960, 5008 in the spectra of galaxies and QSOs have been used to study the variation of $\alpha$ over time.
\cite{Bahcall1967em} analyzed the \oiii doublet in five QSOs at $0.17<z<0.26$ and found no variation in $\alpha$ as they measured $\dalpha=\frac{\alpha(z)}{\alpha(0)}-1=(1\pm2)\times10^{-3}$, where $\alpha(z)$ and $\alpha(0)$ are the $\alpha$ values at redshift $z$ and in laboratory.
The \oiii doublet method was sidelined for decades and returned in the 2000s when the SDSS project produced a large sample of QSO spectra with resolution $R\approx2000$.
The final precision can be greatly improved by averaging the results of the measurements from a large number of QSOs.
\cite{Bahcall2004} measured $\dalpha=(0.7\pm1.4)\times10^{-4}$ using \oiii doublet in spectra of 42 QSOs at $0.16<z<0.80$ from the SDSS Early Data Release.
\cite{Gutierrez2010} measured $\dalpha=(2.4\pm2.5)\times10^{-5}$ using 1568 QSOs at $z<0.8$ from SDSS Data Release 6.
\cite{Rahmani2014} measured $\dalpha=(-2.1\pm1.6)\times10^{-5}$ using 2347 QSOs at $0.02<z<0.74$ from SDSS Data Release 7.
\cite{Albareti2015} measured $\dalpha=(0.9\pm1.8)\times10^{-5}$ using 13,175 QSOs at $z<1$ from SDSS Data Release 12.
In addition to the \oiii doublet, other doublet emission lines had been attempted, such as [Ne III] $\lambda\lambda$3870,3969, to study the $\alpha$ variation at $1<z<1.5$.
However, \cite{Albareti2015} demonstrated that the accuracy is limited to be worse than $10^{-3}$ due to systematic errors caused by the contamination from H$\epsilon$ to [Ne III] $\lambda$3969.

At $z>1$, when the \oiii doublet is redshifted beyond the wavelength ranges of optical spectrometers, the studies on the variation of $\alpha$ using astronomical spectroscopy can be carried out using alkali doublet absorption lines in the ultraviolet spectra of QSOs \citep[e.g.][]{Bahcall1967ab,Murphy2001}.
With a principle similar to the \oiii doublet method, measurements of $\dalpha$ involving this alkali-doublet method have reached a precision of $\sim10^{-5}$, slightly better than the \oiii doublet method and did not detect variation in $\alpha$ either.

The most precise astronomical limits on $\alpha$ variation till now were achieved using many-multiplet method \citep[e.g.,][]{Dzuba1999, Webb2001}.
In this method, the wave number of a spectra line, in producing which the fine-structure splitting of an energy level is involved, can be expressed as $\omega_z=\omega_0+q_1 x+q_2 y$ \citep[e.g.,][]{Dzuba1999}, where $\omega_z$ and $\omega_0$ are the wave numbers in vacuum at redshift $z$ and at today, respectively, and $x=(\frac{\alpha_z}{\alpha_0})^2-1$, $y=(\frac{\alpha_z}{\alpha_0})^4-1$.
The parameters $q_1$ and $q_2$ can be computed using relativistic many-body perturbation and experimental data, while their values for light elements and heavy elements can differ by a factor of tens to thousands. This great diversity of the values of $q_1$ and $q_2$ can amplify possible variation of $\alpha$ and then reveal it in the differences between the wave numbers of the spectral lines of different atoms and ions.
Multiple absorption lines in the damped Ly$\alpha$ absorption systems in QSOs' spectra, including Mg II, Fe II, and others, were used.
In practice, the absorption lines generally have complex profiles, and therefore, a fitting technique involving many Voigt profiles is used.
All absorption lines in an absorption system are fitted simultaneously, in which every line is decomposed into multiple Voigt components, and the number of free parameters is minimized by linking the physically related absorption components of different lines.
Some early works claimed to have found evidence of significant differences between the past and present values of $\alpha$ \citep[e.g.,][]{Webb2001,Murphy2003}.
However, subsequent works \citep[e.g.,][]{Murphy2007,Griest2010,Whitmore2015} indicated that the early results were affected by wavelength distortion.
After taking the wavelength distortion into account, later works, including those using QSO spectra with resolutions of $40,000\sim60,000$ aiming at several to hundreds of absorbers at $0.2<z<4$ \citep[e.g.,][]{King2012,Evans2014,Songaila2014,Murphy2017} and those using spectra of bright QSO HE 0515-4414 with resolutions up to 145,000 \citep[e.g.,][]{Kotus2017,Milakovic2021,Murphy2022qso}, measured $\dalpha$ with precisions of $(0.8\sim4)\times10^{-6}$ and did not detect variation in $\alpha$ over time.
Recently, \cite{Murphy2022star} applied the many-multiple method to the absorption spectra of neighbouring stars within 50 pc, finding $\dalpha$ consistent with 0 with a precision of $1.2\times10^{-8}$.

Though having archived great precision, possible problems lurking in works using the many-multiplet method have been pointed out \citep[see a review by][]{Webb2022}, which may emerge from the techniques for correcting wavelength distortion \citep[e.g.,][]{Dumont2017}, the physically assumed linking that the Voigt fitting technique relies on \citep[e.g.,][]{Levshakov2004}, the technical details of the fitting process \citep[e.g.,][]{Bainbridge2017,Lee2023}, the unconsidered systematic errors \citep[e.g.,][]{Lee2021}, and other.
These problems can lead to biases in the best-fitting $\dalpha$ values or to underestimations of their errors, and some of the problems lack clear solutions despite much effort.
Anyhow, the results of the many-multiplet method need to be tested with entirely different methods.
The \oiii doublet method relies on fewer assumptions and suffers from fewer systematic errors.
No assumptions on chemical composition, ionization state, and distribution of energy levels are required, because the \oiii doublet lines originate in the downward transitions from the same upper level of the same ion.
Also, there is no need to decompose one \oiii emission line into multiple components by assuming that it originates from multiple clouds, as was usually the routine in the study of absorption lines, for the \oiii doublet lines must have the same profile.
The \oiii doublet method is more tolerant of the wavelength distortion because of the small wavelength range used in the measurement (doublet line interval $47.9(1+z)$ \AA).
At present, although the precision of the $\dalpha$ measured by the \oiii doublet method is much lower than that measured by the many-multiplet method, improvement in the future can be achieved by using more spectra of QSOs and starburst galaxies with better quality.
In addition, current measurements based on absorption lines were limited to $z<7.1$ \citep{Wilczynska2020}.
Studying the $\alpha$ variation using absorption lines at higher redshifts is challenging because QSOs are extremely rare in the early universe, and the normal galaxies' continua are too weak.
Fortunately, many starburst galaxies have been discovered in the early universe, and their \oiii doublet or other doublet emission lines can be used.

In studies of $\alpha$ variation using the \oiii doublet method, works with large sample sizes ($N>20$) have used only spectra of QSOs, not starburst galaxies.
This may be because QSOs have higher \oiii luminosities, or it may be related to the observational strategy of the SDSS project.
In reality, starburst galaxies far outnumber QSOs in the universe.
They have narrower \oiii emission lines than QSOs, which is advantageous for improving the accuracy of $\dalpha$ measurement and may compensate for their disadvantage of lower \oiii luminosity.
If this can be proved, the number of available spectra could be significantly increased by including starburst galaxies in studies of $\alpha$ variation, and the precision of the final results improved.
In addition, previous works only used optical spectra (wavelength $<1$ $\mu$m), and hence the $\dalpha$ measurements were limited to $z<1$.
Applying this method to $z>1$ demands infrared spectroscopy.

The XShooter\citep{Vernet2011} is an intermediate-resolution echelle spectrometer mounted on UT2 of the Very Large Telescope (VLT) since 2009.
Entering celestial radiation is split into three arms optimized for the UVB,VIS and NIR wavelength ranges by dichroic mirrors. Each arm has an echelle and a detector, and the wavelength ranges of the UVB, VIS and NIR arms are 299--556, 534--1020 and 994--2478 nm, respectively.
By splicing the spectra from the three arms, a continuous wavelength coverage from 300 to 2470 nm can be achieved.
The XShooter can work under the long-slit or the integral field unit modes.
For the long-slit mode, slits with widths of 0.4 to 1.5$\arcsec$ are available in the NIR arm.
The slits with widths of 0.6, 0.9 and 1.2 $\arcsec$ are often used, producing NIR spectra in resolutions of 8030, 5570 and 4290, respectively.
After an observation finishes on the Xshooter, the raw data will be processed automatically by the pipeline \citep{Modigliani2010}, giving the extracted spectra of the targets.
Overall, the XShooter has a unique wavelength coverage of up to 2.47$\mu$m with a moderate spectra resolution, and it is highly sensitive due to the large aperture of the VLT and the high efficiency of the spectrometer, not to mention the pipeline makes accurate and reliable wavelength calibration.
Thus, spectra taken by the XShooter are suitable for studies of the variation of $\alpha$ via the \oiii doublet method at $z > 1$.

In this work, we measured the $\alpha$ variation by the \oiii doublet method using XShooter spectra of a sample of Lyman-$\alpha$ emitters (LAEs) and QSOs.
LAEs are distant galaxies selected by their strong Lyman-$\alpha$ emission lines, among which most are starburst galaxies, and few are obscured active galactic nuclei.
We achieved a measurement at $1.1<z<3.7$ for the first time with this method.
In addition, we found that $\alpha$ variation can be better constrained by using LAE spectra than QSO spectra.
Throughout the paper, we assume a $\Lambda$CDM cosmology and use the cosmological parameters obtained by \cite{Bennett2014}, which are $H_0=69.6$ km s$^{-1}$ Mpc$^{-1}$, $\Omega_M=0.286$, and $\Omega_\Lambda=0.714$.

\section{Data Reduction and Sample Selection}

\subsection{Data reduction}

We collected information on the normal, the large, the small, the ToO and the DDT XShooter programs in period 84 $-$ period 104 between 2009 and 2020.
Projects related to LAEs and QSOs at $1<z<4$ and observed in the long-slit mode were selected.
The redshifts of the included targets were then determined roughly by inspecting their optical/NIR spectra visually.
Although these rough redshifts were only used for selecting targets,
their accuracy was guaranteed later by the more precise redshifts obtained in further analysis by fitting the spectra.
The uncertainties of the rough redshifts are on the 0.01 level, which is sufficient for selecting reliable targets at this stage.
Targets with redshifts between
1.07 and 3.77 were selected, ensuring the NIR spectra cover the wavelength range of 4800--5200 \AA\ in the rest frame.

We obtained the two-dimensional (2D) and one-dimensional (1D) NIR spectral data of the selected targets from the ESO
database\footnote{The spectral data products query form, http://archive.eso.org/wdb/wdb/adp/phase3\_spectral/form },
which had been processed by the XShooter pipeline (version is related to the observation time).
These data are referred to as pip-2D and pip-1D hereafter.
The pip-2D data underwent processes including wavelength calibration, CCD cleaning, target tracking and straightening, combining different exposures, removing cosmic rays, and merging spectra from different orders.
They had also been interpolated to a wavelength grid with a fixed interval of 0.6 \AA.
Based on the pip-2D data, the pip-1D data underwent spectral extraction and flux calibration.
Instead of using the pip-1D data directly, we used the pip-2D data to extract the final 1D spectrum.
In this section, we briefly introduce the typical extraction approach and display more details and the treatment of some exceptional cases in Appendix A.

Dither mode was used in most observations, and under this mode, a target will experience several consecutive exposures.
The pipeline combines the data from the group of multiple exposures of a target, leaving three parallel images of the target in its pip-2D data, in which one image in the middle has positive flux, and the other two have negative fluxes.
The extraction of such pip-2D data of a target is as follows.
First, we determined the aperture for extraction by fitting the brightness profiles of the three images at the spatial direction of the 2D data.
Second, we extracted the spectra of the three images and combined them.
Third, we made flux calibration and telluric correction for the combined spectrum.
Finally, we identified the unreliable pixels and assigned marking masks for them.
Pixels seriously affected by sky emission lines (SELs) or telluric absorption lines (TALs) were masked in this step, and Figure 1 shows the examples.
In addition, if a target has more than one group of exposures in a project and hence several spectra were extracted, we would combine these extracted spectra into one.

We extracted 95 LAE and 601 QSO spectra (including obscured QSOs).
Note that if one target was observed in different projects, we extracted a spectrum for each project.

\begin{figure}
\centering{
  \includegraphics[scale=0.8]{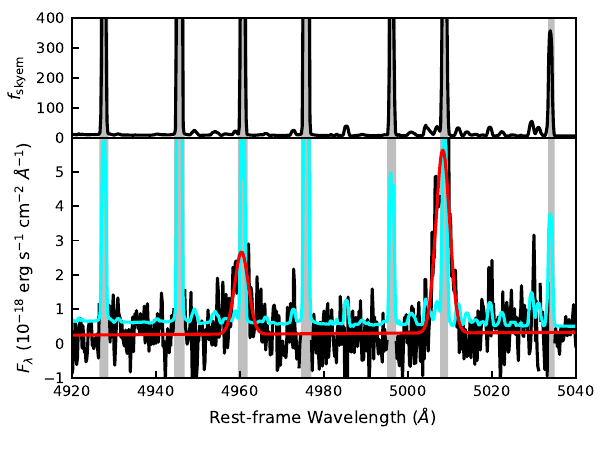}
  \includegraphics[scale=0.8]{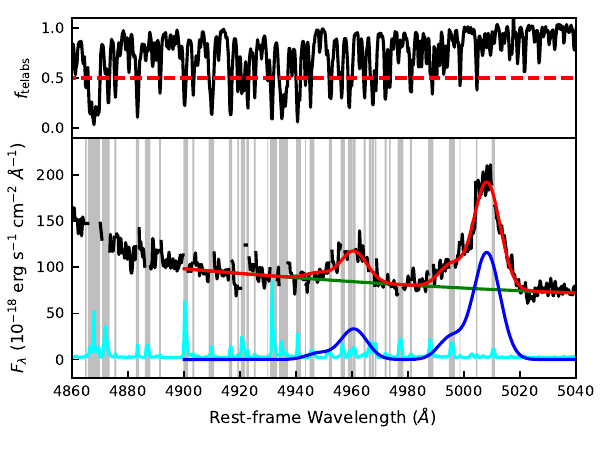}
  \caption{{\bf Left}: Examples of masks caused by SELs.
  We identified strong lines in the sky emission spectrum and labelled the pixels strongly affected by them in grey.\
  {\bf Right}: Examples of masks caused by TALs.
  We show the normalized telluric absorption spectrum (1 for unabsorbed, 0 for fully absorbed).
  Pixels with a value less than 0.5 (red dotted line) on the normalized absorption spectrum are masked (grey).}}
  \label{fig1}
\end{figure}

\subsection{Spectral Analysis}

Aimed to build a sample with spectra that have strong and clean \oiii doublet lines,
we fit the \oiii doublet lines and their adjacent regions in each spectrum for the present.
The fitting model includes one component for the continuum and one for the \oiii doublet.
The former component is a linear function for an LAE and a second-order polynomial for a QSO.
We did not set a separate component for the H$\beta$ broad emission line in a QSO spectrum because it has been included in the second-order polynomial.
For the latter component, a sum of multiple Gaussian components was used for each \oiii line.
The $\lambda$4960 and $\lambda$5008 lines contain the same number of Gaussian components, and for each Gaussian component, we set:
\begin{equation}
\left\{
     \begin{array}{lr}
       { w_{i,5008}=w_{i,4960}\times(1+\eta) } \\
       { \sigma_{i,5008}=\sigma_{i,4960}\times(1+\eta) } \\
       { f_{i,5008}=f_{i,4960}\times A }
     \end{array} \right.
\end{equation}
where $w_{i,5008}$, $\sigma_{i,5008}$ and $f_{i,5008}$ are the central wavelength, standard deviation and flux of the $i$th Gaussian component of the $\lambda$5008 line, and those with a subscript 4960 are the corresponding parameters of the $\lambda$4960 line.
The definitions of $\eta$ and $A$ followed \cite{Bahcall2004}, where $1+\eta$ is the ratio of the wavelengths of the doublet, and $A$ is the ratio of the fluxes of the doublet.
We fixed $\eta$ as the theoretical value $\eta_0$ in this stage to distinguish the doublet lines when blended or avoid misidentification of the $\lambda$4960 line when weak.
We adopted the vacuum wavelengths of 4960.295 and 5008.240 \AA\ for the doublet, which were taken from the NIST Atomic Spectra Database\footnote{https://www.nist.gov/pml/atomic-spectra-database}, and hence $\eta_0=0.00966576$.
For spectra with $>10\sigma$ detections of both doublet lines, fixing $\eta$ to $\eta_0$ generally leads to an increase of the signal-to-noise ratios (SNR) of the $\lambda$4960 by $<$20\%, and an increase of that of the $\lambda$5008 by $<$5\%.
Here an SNR was calculated as the flux of a line divided by its error.

The rest-wavelength ranges for fitting are different for LAEs and QSOs.
In most cases, the rest-wavelength range for LAEs is 4900--5100 \AA\ and 4900--5140 \AA\ for QSOs.
When fitting some QSOs' spectra, the result contains Gaussian components with FWHMs of several $10^4$ km s$^{-1}$, which lack astronomical interpretations and are actually caused by a miscalculation of the continuum.
In these cases, we adjusted the red end of the range to 5160, 5180 or 5200 \AA\ to define the continuum component better.
For QSOs with broad and blueshifted \oiii emission lines, such as the obscured QSOs with strong \oiii outflows in the sample of \cite{Zakamska2016}, we adjusted the blue end of the range to 4880 \AA\ to decompose the continuum and the $\lambda$4960 line effectively.
For any spectrum, pixels with masks were excluded from the fitting.

We fit the spectra by minimizing $\chi^2$.
First we tried one Gaussian component for each line, and then gradually added more Gaussian components.
One component more will lower the $\chi^2$ and the degree of freedom by 3 at the same time.
Supposing the fitting were improved after adding a component, we calculated the confidence probability using the F test:
\begin{equation}
F = \frac{\Delta\chi^2/\chi^2}{\Delta dof/dof} ,
\end{equation}
where $\chi^2$ and $dof$ are the chi-square and the degree of freedom after adding the component, $\Delta\chi^2$ and $\Delta dof$ are the decreased amounts, and $\Delta dof$ is 3.
We calculated the probability $p(F, dof, \Delta dof)$ corresponding to this $F$ value.
If $p$ is greater than 99.99\%, we would add this component, and vice versa.
This threshold of $p$ corresponds to $\Delta\chi^2$ of 20--30 for most spectra ($dof$ is 1000--1500).
Figure 2 shows an example of how we gradually added the Gaussian components.
All the doublets in the spectra can be fitted with no more than 4 components, while when 4 components have been used, adding one more only improves the fitting slightly but raise difficulty in finding the best solution.

\begin{figure}
\centering{
  \includegraphics[scale=0.8]{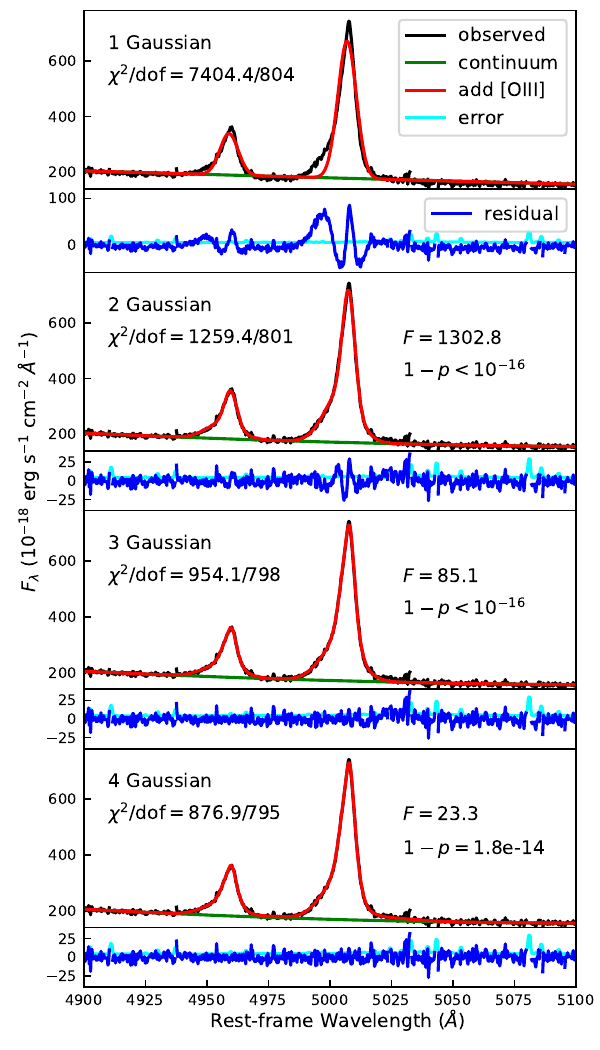}
  \caption{
  Example of the multiple Gaussian fitting.
  From top to bottom panels, we show the models containing 1--4 Gaussian components for \oiii doublet.
  In each panel, we display the observed spectrum (black), the continuum model (green), the sum of continuum and emission line models (red), the residual (blue), and the observed errors (cyan).
  Note that we only show spectra and model in 4900--5100 \AA\ while the wavelength range of fitting is 4900 to 5140 \AA.
  }}
  \label{fig2}
\end{figure}

\cite{Bahcall2004} indicated that the SDSS pipeline overestimated the errors for lower flux levels(see their section 4.4).
This situation also happened to the Xshooter pipeline.
Therefore we corrected the error following \cite{Bahcall2004}.
If the fitting's reduced chi-square ($\chi^2/dof$) is less than 1 when the original error is used, we multiplied the error of all pixels by a correction factor (equal to the square root of the reduced chi-square) and redid the fit.
We corrected the error for 43\% of the spectra, and the correction factors in our final sample are 0.6--1.0.
This correction decreases the errors of model parameters and increases the SNRs of \oiii doublet lines.
After correction, the SNRs are more consistent with visual estimates (e.g., 3--5 for a weak detection or $>7$ for a clear detection).

\subsection{Sample Selection}

Our criteria for sample selection include the following aspects.

First, we required the SNRs of both the $\lambda$4960 and $\lambda$5008 lines to be above 10.

Second, we required that the \oiii doublet in a spectrum hardly bears any SELs, TALs, cosmic rays or bad CCD pixels.
Although the pixels affected by these factors had been masked in Section 2.1 and were excluded from spectral analysis, these factors can still cause problems because automated programs may fail to mask all affected pixels when the effects are severe.
In addition, for an emission line with half the pixels masked, including those near the peak, setting $\eta$ as a free parameter (as we will do in Section 3) may lead to mistaken identification of the peak.
We calculated a mask index $I_{\rm mask}$ for each line to describe these effects quantitatively.
As shown in Figure 1, this index was calculated as the sum of the fluxes of the masked pixels (grey region) divided by the total flux of the line.
After visual inspections of the effects mentioned above, we required that the sum of $I_{\rm mask}$ of $\lambda$4960 and $\lambda$5008 lines be below 0.5.
The LAE spectra excluded by this criterion were mainly due to SELs, while the QSO spectra excluded were mainly due to TALs.

Third, we excluded QSOs with \oiii doublet lines severely blended.
If so, the decomposition of the doublet may be wrong by setting $A$ as free parameters if that by fixing $A$ at theoretical value treated as correct.
A blending index $I_{\rm blend}$ was calculated to quantify the blending, as is shown in Figure 3(b).
On the multiple Gaussian models of the \oiii doublet lines, we measured the flux of the highest point near 4960 \AA\ as $f_1$ and the flux of the lowest point in the range between 4960--5008 \AA\ as $f_2$. $I_{\rm blend}$ was set to be the ratio between $f_2$ and $f_1$.
According to visual inspections, to ensure the doublet lines be properly decomposed, we required that $I_{\rm blend}$ is below 0.2.
The QSO spectra excluded by this criterion all have broad and blueshifted \oiii emission lines.
The decomposition by fixing $A$ at theoretical value show that the \oiii FWHMs of these excluded QSOs are 800--4500 km s$^{-1}$, significantly greater than the FWHMs of the final sample, which has a median value of 610 km s$^{-1}$.

Finally, we excluded QSOs with strong Fe II bumps.
The Fe II bump between 4900 and 5050 \AA\ may lead to a severe systematic error when fitting the continuum with a simple second-order polynomial.
To measure the intensity of the Fe II bump, we calculated a Fe II index $I_{\rm FeII}$ as follows.
As shown in Figure 3(a), we measured $I_{4590}$ and $I_{5250}$, which represent the intensity of the Fe II bumps around 4590 \AA\ (Fe II $\lambda$4590) and around 5100--5400 \AA\ (Fe II $\lambda$5250).
We inferred the continuum below the Fe II bumps using the spectra in the green shade in Figure 3(a): the wavelength ranges for fitting the continuum are 4420-4460 \AA\ and 4720--4760 \AA\ for Fe II $\lambda$4590, and 5060--5100 \AA\ and 5400--5440 \AA\ for Fe II $\lambda$5250; the model used was a power law; the best-fitting model for the continuum is expressed as $f_{\rm con}(\lambda)$, as is shown by the blue line in Figure 3(a).
We fit the continuum-subtracted spectra in 4460--4720 \AA\ and 5100--5400 \AA\ using cubic spline functions whose nodes have a uniform interval of 20 \AA, and the fitting result is expressed as $f_{\rm FeII}(\lambda)$.
Hence we calculated:
\begin{equation}
I_{4590} = \frac{ \int_{4460}^{4720} f_{\rm FeII}(\lambda) {\rm d}\lambda }{ f_{\rm con}(4590 \AA ) },\
I_{5250} = \frac{ \int_{5100}^{5400} f_{\rm FeII}(\lambda) {\rm d}\lambda }{ f_{\rm con}(5250 \AA ) }
\end{equation}
Note that for most QSOs, only one of $I_{4590}$ and $I_{5250}$ can be reliably measured because of the telluric absorption bands.
For uniformity, if $I_{5250}$ can be reliably measured, then we set $I_{\rm FeII}=I_{5250}$.
And if not, we set $I_{\rm FeII}=1.3\times I_{4590}$.
The coefficient 1.3 here was chosen to be the ratio between $I_{5250}$ (16 \AA) and $I_{4590}$ (12 \AA) measured on the Xshooter QSO composite spectrum \citep{Selsing2016}.
We excluded QSOs that meets the following criteria: $I_{\rm FeII}>10 \AA$ and $I_{\rm FeII}>0.5\times EW_{\rm [OIII]}$, where $EW_{\rm [OIII]}$ is the equivalent width of \oiii $\lambda$5008.

\begin{figure}
\centering{
  \includegraphics[scale=0.8]{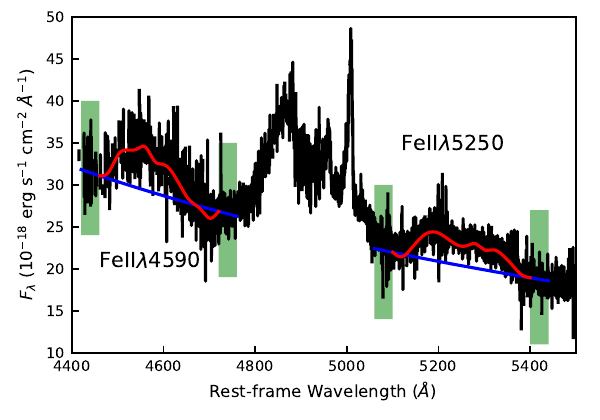}
  \includegraphics[scale=0.8]{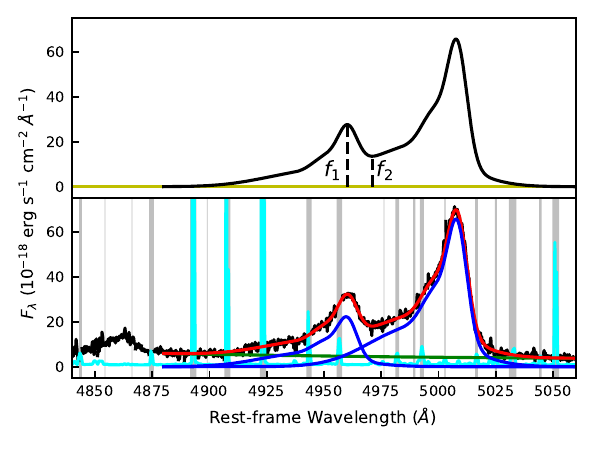}
  \caption{
  {\bf (a)}: An example of measuring $I_{\rm FeII}$.
  The green shade shows the continuum.
  The continuum model is blue, and the wavelength ranges used for fitting it are green shades.
  The red line shows the sum of the continuum and Fe II models.
  {\bf (b)}: An example of measuring $I_{\rm blend}$.
  The black line in the upper panel is the sum of $\lambda$5008 and $\lambda$4960 models.
  }}
  \label{fig3}
\end{figure}

The final sample passing these criteria contains 86 spectra, including 40 spectra of 32 LAEs and 46 spectra of 45 QSOs.
The observing information of these 86 spectra is listed in Table 1 (the LAE subsample) and Table 2 (the QSO subsample).
Six of the LAE spectra come from a galaxy at $z=2.37$ named ``Sunburst arc'' \citep[SA,][]{Dahle2016,Rivera-Thorsen2017}, which is highly magnified by a gravitational lens (magnification $>20$).
We refer to these six spectra as SA1 to SA6.
LAE J0332-2746 has three spectra, J0217-0502 has two spectra, and each of the other 29 LAEs has only one spectrum.
QSO J1313-2716 has two spectra, and each of the other 44 QSOs has one spectrum.

\begin{table*}
\caption{Information of the 40 LAE spectra}
\scriptsize
\begin{tabular}{ccccccccccc}
\hline
ProgID & ObjID & Ra & Dec & z & $t_{\rm exp}$ & $R$ & log $L_{5008}$ & FWHM & SNR$_{4960}$ & SNR$_{5008}$\\
\hline
 084.A-0303(C)  & J1000+0201  & 10:00:49.17 & +02:01:19.4 & 2.246 & 3600  & 4287 & 42.48 & 95   & 10.89  & 30.23 \\
 088.A-0672(A)  & J0332-2759  & 03:32:49.44 & -27:59:51.6 & 2.205 & 9600  & 5573 & 42.40 & 112  & 10.02  & 37.63 \\
                & J0332-2800  & 03:32:32.41 & -28:00:51.8 & 2.172 & 9600  & 5573 & 43.05 & 182  & 33.03  & 65.07 \\
 091.A-0413(A)  & J2346+1247  & 23:46:09.17 & +12:47:54.5 & 2.327 & 4640  & 8014 & 42.24 & 176  & 10.20  & 16.68 \\
                & J2346+1249  & 23:46:29.41 & +12:49:42.9 & 2.173 & 9280  & 8014 & 42.41 & 138  & 23.12  & 46.13 \\
 093.A-0882(A)  & J0217-0511  & 02:17:47.27 & -05:11:44.1 & 2.243 & 7200  & 4287 & 42.40 & 131  & 23.44  & 50.26 \\
                & J0217-0512  & 02:17:43.41 & -05:12:45.5 & 2.183 & 3600  & 5573 & 42.76 & 115  & 18.38  & 46.60 \\
                & J0332-2745  & 03:32:24.53 & -27:45:34.9 & 2.311 & 7200  & 4287 & 42.88 & 131  & 41.96  & 87.17 \\
                & J0332-2746  & 03:32:36.85 & -27:46:18.3 & 2.226 & 3600  & 5573 & 42.68 & 164  & 19.74  & 43.15 \\
                & J2358-1014  & 23:58:21.33 & -10:14:30.8 & 2.261 & 10800 & 4287 & 42.68 & 133  & 30.13  & 37.91 \\
 096.A-0348(A)  & J0217-0514  & 02:17:03.97 & -05:14:16.2 & 2.185 & 7200  & 4287 & 42.69 & 132  & 11.82  & 41.86 \\
                & J0217-0515  & 02:17:43.60 & -05:15:27.1 & 2.065 & 3600  & 4287 & 42.56 & 143  & 11.10  & 25.66 \\
                & J0332-2741  & 03:32:38.91 & -27:41:37.5 & 2.312 & 7200  & 4287 & 42.74 & 207  & 22.31  & 53.97 \\
 097.A-0153(A)  & J0145-0946  & 01:45:16.78 & -09:46:04.5 & 2.357 & 7200  & 5573 & 42.93 & 221  & 42.37  & 103.25\\
                & J0209-0005A & 02:09:49.27 & -00:05:33.0 & 2.167 & 3600  & 5573 & 42.65 & 133  & 22.69  & 32.17 \\
                & J0209-0005B & 02:09:43.19 & -00:05:51.9 & 2.187 & 7200  & 5573 & 42.79 & 225  & 23.87  & 37.23 \\
 099.A-0254(A)  & J0217-0502  & 02:17:46.02 & -05:02:58.6 & 2.209 & 8960  & 4287 & 42.45 & 140  & 11.31  & 19.81 \\
 099.A-0758(A)  & J0332-2746  & 03:32:35.39 & -27:46:15.1 & 2.171 & 13200 & 5573 & 42.53 & 119  & 18.64  & 32.16 \\
                & J0332-2749  & 03:32:17.95 & -27:49:42.7 & 2.033 & 10800 & 5573 & 42.60 & 148  & 20.27  & 68.82 \\
 0101.B-0262(A) & J0014-3024  & 00:14:20.71 & -30:24:30.8 & 2.016 & 4800  & 5573 & 42.34 & 109  & 10.31  & 23.39 \\
                & J0120-5143  & 01:20:42.35 & -51:43:52.0 & 1.295 & 4800  & 5573 & 42.10 & 103  & 9.99   & 22.83 \\
                & J0131-1335  & 01:31:51.96 & -13:35:38.1 & 2.467 & 4800  & 5573 & 42.63 & 150  & 18.91  & 42.20 \\
                & J2129+0006  & 21:29:39.55 & +00:06:48.9 & 2.405 & 4800  & 5573 & 42.38 & 163  & 16.26  & 31.20 \\
                & J2129-0741  & 21:29:21.92 & -07:41:02.1 & 2.274 & 4800  & 5573 & 42.71 & 188  & 19.86  & 59.34 \\
                & J2140-2339  & 21:40:19.08 & -23:39:30.9 & 2.486 & 4800  & 5573 & 42.95 & 148  & 28.84  & 44.96 \\
                & J2248-4431  & 22:48:37.00 & -44:31:18.5 & 2.065 & 4800  & 5573 & 42.49 & 123  & 11.99  & 30.04 \\
 0101.B-0779(A) & J1000+0236  & 10:00:37.34 & +02:36:45.5 & 3.423 & 7200  & 5573 & 43.17 & 164  & 20.45  & 40.89 \\
 0102.A-0391(A) & J0332-2745  & 03:32:03.18 & -27:45:17.7 & 3.211 & 14400 & 5573 & 43.01 & 129  & 20.34  & 51.12 \\
 0102.A-0652(A) & J0217-0502  & 02:17:46.19 & -05:02:55.6 & 2.209 & 8800  & 5573 & 42.71 & 118  & 25.07  & 45.14 \\
 0103.A-0688(C) & SA1         & 15:50:04.55 & -78:10:57.1 & 2.369 & 2700  & 5573 & 44.37 & 124  & 163.07 & 266.60\\
                & SA2         & 15:50:04.55 & -78:10:57.1 & 2.369 & 2700  & 5573 & 44.29 & 114  & 187.94 & 467.47\\
                & SA3         & 15:50:00.02 & -78:11:12.9 & 2.369 & 2700  & 8031 & 44.12 & 114  & 113.17 & 167.20\\
                & SA4         & 15:50:00.02 & -78:11:12.9 & 2.369 & 2700  & 8031 & 43.86 & 136  & 94.52  & 259.79\\
                & SA5         & 15:49:59.05 & -78:11:12.4 & 2.369 & 1800  & 8031 & 43.96 & 116  & 114.40 & 209.85\\
                & SA6         & 15:49:59.05 & -78:11:12.4 & 2.369 & 1800  & 8031 & 43.68 & 151  & 71.30  & 184.56\\
 0103.B-0446(A) & J0224-0449  & 02:24:18.99 & -04:49:12.3 & 2.550 & 14400 & 5573 & 42.68 & 123  & 32.15  & 64.78 \\
                & J0332-2752  & 03:32:12.98 & -27:52:34.7 & 2.566 & 1800  & 5573 & 42.99 & 365  & 21.42  & 43.28 \\
                & J1001+0206  & 10:00:08.86 & +02:06:39.7 & 2.435 & 1800  & 5573 & 42.73 & 489  & 10.62  & 19.49 \\
 0104.A-0236(A) & J0332-2746  & 03:32:35.31 & -27:46:18.1 & 2.171 & 21600 & 5573 & 42.25 & 127  & 19.13  & 57.12 \\
 0104.A-0236(B) & J0332-2746  & 03:32:35.32 & -27:46:17.9 & 2.171 & 12600 & 5573 & 42.29 & 132  & 14.89  & 48.34 \\
\hline
\end{tabular}
\label{tab1}
\end{table*}

\begin{table*}
\caption{Information of the 46 QSO spectra}
\scriptsize
\begin{tabular}{ccccccccccc}
\hline
ProgID & ObjID & Ra & Dec & z & $t_{\rm exp}$ & $R$ & log $L_{5008}$ & FWHM & SNR$_{4960}$ & SNR$_{5008}$\\
\hline
 086.B-0320(A)  & J1242+0232  & 12:42:20.04 & +02:32:55.6 & 2.221 & 3000  & 4287 & 43.71 & 842  & 11.62  & 18.53 \\
 087.A-0610(A)  & J0914+0109  & 09:14:32.07 & +01:09:10.0 & 2.139 & 2400  & 8031 & 43.27 & 792  & 14.79  & 34.64 \\
 087.B-0229(A)  & J1512+1119  & 15:12:49.39 & +11:19:27.3 & 2.100 & 8700  & 5573 & 44.46 & 560  & 122.60 & 142.96\\
 088.B-1034(A)  & J0148+0003  & 01:48:12.79 & +00:03:21.6 & 1.481 & 2880  & 4287 & 43.42 & 550  & 10.29  & 25.29 \\
                & J0240-0758  & 02:40:28.97 & -07:58:45.8 & 1.533 & 5760  & 4287 & 43.40 & 335  & 16.65  & 24.15 \\
                & J0323-0029  & 03:23:49.63 & -00:29:51.3 & 1.625 & 2880  & 4287 & 43.71 & 483  & 16.71  & 26.16 \\
                & J0842+0151  & 08:42:40.68 & +01:51:31.4 & 1.496 & 1440  & 4287 & 43.51 & 972  & 13.54  & 46.10 \\
                & J1005+0245  & 10:05:13.74 & +02:45:08.0 & 1.484 & 2880  & 4287 & 43.76 & 577  & 21.83  & 30.65 \\
 089.B-0275(A)  & J1226-0006  & 12:26:07.94 & -00:06:03.1 & 1.126 & 2400  & 5567 & 43.01 & 610  & 25.91  & 47.80 \\
 089.B-0936(A)  & J0904-2552  & 09:04:52.43 & -25:52:51.0 & 1.644 & 1260  & 5567 & 43.29 & 385  & 12.59  & 20.73 \\
                & J1107-4449  & 11:07:08.47 & -44:49:07.8 & 1.598 & 960   & 5567 & 43.52 & 485  & 14.00  & 41.30 \\
 089.B-0951(A)  & J0958+0251  & 09:58:52.18 & +02:51:52.8 & 1.408 & 3780  & 5567 & 43.18 & 395  & 22.65  & 38.30 \\
                & J1313-2716  & 13:13:47.30 & -27:16:46.1 & 2.198 & 760   & 5567 & 44.16 & 822  & 23.80  & 35.45 \\
 090.A-0830(A)  & J1002+0137  & 10:02:11.41 & +01:37:08.0 & 1.591 & 3600  & 5567 & 42.86 & 833  & 11.99  & 31.60 \\
                & J1003+0209  & 10:03:08.99 & +02:09:03.2 & 1.472 & 3600  & 5567 & 42.99 & 1219 & 10.88  & 34.57 \\
 090.B-0424(B)  & J0811+1720  & 08:11:14.58 & +17:20:55.2 & 2.332 & 14400 & 5573 & 44.10 & 643  & 72.58  & 133.05\\
 091.B-0900(A)  & J1313-2716  & 13:13:47.17 & -27:16:50.1 & 2.200 & 1030  & 5567 & 44.10 & 766  & 22.14  & 39.14 \\
                & J1404-0130  & 14:04:45.89 & -01:30:24.6 & 2.515 & 1860  & 5567 & 43.71 & 373  & 46.12  & 113.68\\
                & J1544+0407  & 15:44:59.42 & +04:07:43.9 & 2.182 & 1860  & 5567 & 43.99 & 695  & 57.54  & 140.20\\
 092.A-0391(A)  & J0209-0438  & 02:09:30.68 & -04:38:30.0 & 1.132 & 2400  & 5573 & 43.94 & 441  & 80.07  & 115.24\\
 092.B-0860(A)  & J1154-0215  & 11:54:32.66 & -02:15:40.6 & 2.181 & 7200  & 5573 & 43.21 & 392  & 23.04  & 59.62 \\
 093.B-0553(A)  & J0940+0034  & 09:40:25.62 & +00:33:58.8 & 2.332 & 1920  & 5573 & 43.44 & 511  & 13.41  & 25.51 \\
                & J1019+0345  & 10:19:25.94 & +03:45:33.9 & 2.212 & 2400  & 5573 & 43.25 & 702  & 12.93  & 31.01 \\
                & J1323+0022  & 13:23:07.77 & +00:22:59.6 & 2.258 & 1920  & 5573 & 43.62 & 438  & 31.80  & 65.65 \\
                & J2330-0216  & 23:30:16.90 & -02:16:45.9 & 2.496 & 1920  & 5573 & 43.46 & 493  & 14.83  & 25.71 \\
 094.B-0111(A)  & J0108+0134  & 01:08:38.66 & +01:34:57.6 & 2.099 & 1800  & 5567 & 43.65 & 560  & 28.85  & 57.16 \\
 095.B-0507(A)  & J0114-0812  & 01:14:20.47 & -08:12:45.5 & 2.104 & 2400  & 4287 & 43.45 & 717  & 21.29  & 42.48 \\
                & J2136-1631  & 21:36:55.62 & -16:31:39.8 & 1.659 & 2400  & 4287 & 43.82 & 397  & 25.70  & 30.77 \\
 098.B-0556(A)  & J0111-3502  & 01:11:43.50 & -35:02:58.3 & 2.405 & 4800  & 5573 & 44.63 & 473  & 240.67 & 399.55\\
                & J0331-3824  & 03:31:06.20 & -38:24:04.9 & 2.435 & 4800  & 5573 & 44.11 & 673  & 99.02  & 211.71\\
 099.A-0018(A)  & J1338+0010  & 13:38:31.46 & +00:10:54.4 & 2.300 & 4800  & 5567 & 43.02 & 486  & 15.63  & 29.17 \\
 099.B-0118(A)  & J0020-1540  & 00:20:25.22 & +15:40:52.5 & 2.020 & 1200  & 5567 & 43.93 & 721  & 42.93  & 76.68 \\
 0101.A-0528(A) & J2124-4948  & 21:24:28.77 & -49:48:07.4 & 2.480 & 1800  & 4287 & 43.57 & 271  & 24.67  & 44.02 \\
 0101.A-0528(B) & J0607-6031  & 06:07:55.01 & -60:31:54.9 & 1.097 & 1800  & 4287 & 43.40 & 481  & 19.57  & 56.62 \\
 0101.B-0739(A) & J1352+1302  & 13:52:18.38 & +13:02:39.7 & 1.600 & 800   & 4287 & 43.76 & 859  & 12.33  & 51.04 \\
                & J1616+0931  & 16:16:39.73 & +09:31:15.7 & 1.468 & 1200  & 4287 & 43.66 & 420  & 25.73  & 72.43 \\
                & J2239+1222  & 22:39:32.01 & +12:22:22.3 & 1.484 & 400   & 4287 & 44.13 & 997  & 20.61  & 57.26 \\
 0102.A-0335(A) & J0259+1635  & 02:59:42.96 & -16:35:41.6 & 2.160 & 4800  & 4287 & 44.64 & 525  & 460.18 & 641.76\\
                & J0630-1201  & 06:30:09.16 & -12:01:18.7 & 3.344 & 4800  & 4287 & 43.96 & 763  & 11.69  & 22.70 \\
                & J1042+1641  & 10:42:22.25 & +16:41:16.7 & 2.519 & 4800  & 4287 & 44.84 & 963  & 287.70 & 300.13\\
 0103.A-0253(A) & J2245-2945  & 22:45:31.61 & -29:45:54.3 & 2.379 & 3200  & 4287 & 43.46 & 341  & 15.60  & 39.16 \\
 189.A-0424(A)  & J1018+0548  & 10:18:18.58 & +05:48:20.8 & 3.512 & 3600  & 5573 & 44.22 & 616  & 17.57  & 37.83 \\
                & J1042+1957  & 10:42:34.03 & +19:57:16.3 & 3.628 & 3600  & 5573 & 44.21 & 1267 & 12.76  & 29.44 \\
                & J1117+1311  & 11:17:01.97 & +13:11:13.1 & 3.623 & 3600  & 5573 & 44.09 & 832  & 12.48  & 28.53 \\
                & J1312+0841  & 13:12:42.94 & +08:41:02.9 & 3.735 & 3600  & 5573 & 44.36 & 859  & 13.90  & 27.94 \\
                & J1621-0042  & 16:21:17.05 & -00:42:52.9 & 3.703 & 3600  & 5573 & 44.44 & 914  & 16.72  & 42.25 \\
\hline
\end{tabular}
\label{tab2}
\end{table*}

For each spectrum, we measured the peak wavelength of the \oiii $\lambda$5008 line and calculated a \oiii redshift as the approximation of the systematic redshift.
For typical QSOs, this would cause an error of $<$0.001 ($<$300 km s$^{-1}$) except for a small fraction of QSOs called  ``blue outliers'' \citep{Marziani2016}, but those blue outliers generally have blended \oiii doublet and would not meet our criteria and hence would not be included in our sample.
We also measured the luminosity of \oiii $\lambda$5008 in each spectrum.
The statistical errors of the luminosities are $<$6\% , corresponding to $<$0.03 dex in logarithm.
Changes in the slit width, the seeing condition and the atmospheric transmittance can cause systematic errors in the luminosities, and we estimated them to be 0.1--0.2 dex based on the multiple observations of some targets.
These redshifts and \oiii luminosities of the spectra are listed in Table 1 and Table 2 and are displayed in Figure 4.
We found that all the targets are distributed in three redshift intervals, $1.09<z<1.66$, $2.01<z<2.57$, and $3.21<z<3.74$, corresponding to the observed wavelengths of the J, H and K bands respectively.
Most of the LAEs are in the redshift range corresponding to the H-band, and this may be caused by the methods with which the LAEs were selected.
The \oiii luminosities of the QSOs are in the range between $10^{42.8}$ and $10^{44.8}$ erg s$^{-1}$, and have a median value of $10^{43.7}$ erg s$^{-1}$.
The \oiii luminosities of the SA spectra are in the range of $10^{43.6-44.4}$ erg s$^{-1}$, and those of other LAEs are in the range between $10^{42.1}$ and $10^{43.2}$ erg s$^{-1}$, and have a median value of $10^{42.7}$ erg s$^{-1}$, which is one order of magnitude lower than that of QSOs.

\begin{figure}
\centering{
  \includegraphics[scale=0.8]{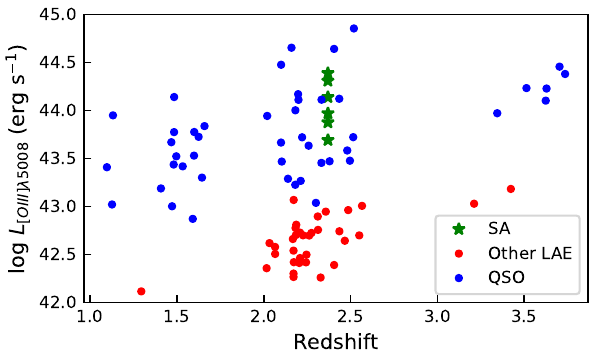}
  \caption{Redshift versus \oiii luminosity for our final sample.}}
  \label{fig4}
\end{figure}

We measured the FWHMs of the \oiii doublet lines in every spectrum of the final sample and showed them in Figure 5 with the SNR of \oiii $\lambda$4960 (SNR$_{4960}$).
These two parameters are critical indicators for the measuring precision of the \oiii wavelengths (see Section 3.3 and Appendix D for analysis and simulation).
In brief, the larger the SNR$_{4960}$ or the smaller the FWHM, the more accurate the measurement of the wavelengths.
The SNR$_{4960}$ of the six SA spectra is 60--200, that of other LAE spectra is 10--50, and that of the QSO spectra is 10--460.
The median values of the SNR$_{4960}$ of the LAE and the QSO subsamples are both around 20.
The FWHM of \oiii of the LAEs is 90--490 km s$^{-1}$ with a median value of 130 km s$^{-1}$, and that of the QSOs is 270--1260 km s$^{-1}$ with a median value of 610 km s$^{-1}$.
The \oiii emission lines of the LAEs are narrower and less luminous than those of the QSOs, while the SNRs of the two subsamples are similar.

\begin{figure}
\centering{
  \includegraphics[scale=0.8]{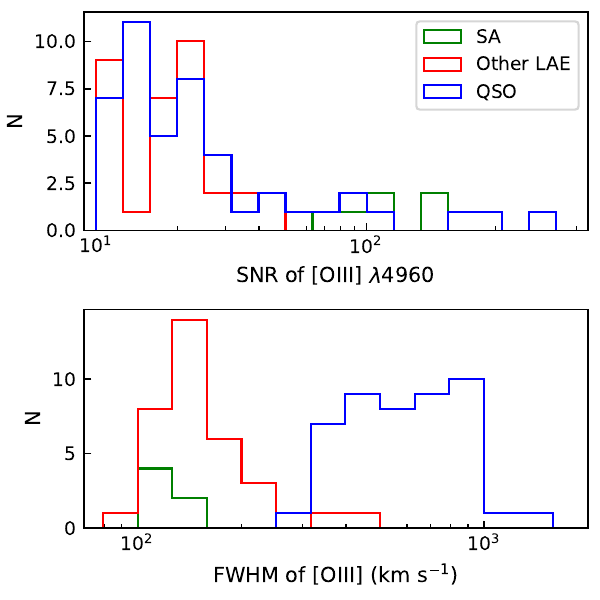}
  \caption{The distributions of SNR$_{4960}$ and \oiii FWHM for the final sample.}}
  \label{fig5}
\end{figure}

\section{Measurements of Emission-line Wavelengths}

Following \cite{Uzan2003}, we calculated the variation in the fine structure constant as:
\begin{equation}
\frac{\Delta\alpha}{\alpha} = \frac{1}{2} \left\{ \frac{[(\lambda_2-\lambda_1)/(\lambda_2+\lambda_1)]_z}{[(\lambda_2-\lambda_1)/(\lambda_2+\lambda_1)]_0} -1
 \right\},
\end{equation}
where $\lambda_1$ and $\lambda_2$ are the wavelengths of $\lambda$4960 and $\lambda$5008 lines, respectively.
Those with a subscript of 0 are the wavelength values at $z=0$ from laboratories, and those with a subscript of $z$ are the observed values of the doublet from a target's spectrum at a redshift of $z$.
Equation (5) can be rewritten as:
\begin{equation}
\frac{\Delta\alpha}{\alpha} = \frac{ \eta-\eta_0 }{ \eta_0(2+\eta) },
\end{equation}
where $\eta \equiv (\lambda_2/\lambda_1-1)_z$, $\eta_0 \equiv (\lambda_2/\lambda_1-1)_0=0.00966576$.
The Taylor expansion of equation (6) around $\eta_0$ gives:
\begin{equation}
\frac{\Delta\alpha}{\alpha} = 51.4802(\eta-\eta_0) - 25.6163(\eta-\eta_0)^2 + o\left( (\eta-\eta_0)^2 \right).
\end{equation}
We will show later that for our sample, $|\eta-\eta_0|$ is less than $3\times10^{-4}$.
In this case, ignoring the quadratic and the high-order terms only results in a relative error of $\Delta\alpha/\alpha$ no more than $\sim10^{-4}$.
Thus, we calculated the change in the fine structure constant and estimated its error with:
\begin{equation}
\frac{\Delta\alpha}{\alpha} \approx 51.4802(\eta-\eta_0).
\end{equation}
In this way, the measurement of $\dalpha$ can be converted into the measurement of $\eta$.

\subsection{Two Methods to Measure $\eta$}

We measured $\eta$ using the following two methods.
The first is called the Multiple-Gaussian (MG) method.
The $\lambda$4960, $\lambda$5008 lines were fitted simultaneously with the model expressed as equation (2).
The fitting is basically the same as that described in section 2.2, and the only difference is that $\eta$ is now a free parameter instead of the fixed value of $\eta_0$.
The wavelength ranges and the number of Gaussian functions are consistent with those used in section 2.2.
We fit the spectra using a Levenberg-Marquardt technique with the MPFIT package.
We set the initial parameters as those obtained in section 2.2.
We took the values of $\eta$ and $A$ yielding the minimum $\chi^2$ (expressed as $\chi_{\rm min}^2$) as the measurement result and calculated the statistical errors $\sigma_{\rm stat}(\eta)$ and $\sigma_{\rm stat}(A)$ according to the range of $\eta$ and $A$ yielding $\chi^2(\eta,A) < \chi_{\rm min}^2 + 1$.
An example of measuring $\eta$ and $A$ using the MG method is shown in the upper row of Figure 6.

The second is called the Profile-Matching (PM) method.
This method is based on the fact that the \oiii doublet originates from the same upper energy level and should have identical line profiles.
The analysis approach is similar to that adopted by \cite{Bahcall2004}.
After subtracting the best-fitting continuum model from the observed spectrum, we obtained the profiles of $\lambda$5008 and $\lambda$4960 emission lines.
We moved the $\lambda$5008 line leftward ($\lambda$ divided by $1+\eta$) and decreased the amplitude ($f_\lambda$ divided by $\frac{A}{1+\eta}$).
Theoretically, at some specific values of $\eta$ and $A$, the adjusted $\lambda$5008 line should be the same with the $\lambda$4960 line, in which case they can be regarded as two samplings from one profile beneath.
So for a pair of values of $\eta$ and $A$, we fit the adjusted $\lambda$5008 line and the $\lambda$4960 line simultaneously with a cubic spline function, from which a $\chi^2$ value was obtained.
With different values of $\eta$ and $A$ yielding different values of $\chi^2$, we obtained a $\chi^2$ surface in the 2D-parameter space.
The $\eta$ and $A$ yielding the minimum $\chi^2$ were taken as the result, and their statistical errors were obtained using $\chi^2(\eta,A) < \chi_{\rm min}^2 + 1$.
In the matching, we used the pixels in the following wavelength ranges:
the range for the $\lambda$5008 is where the flux exceeds 20\% of the peak flux (we use the flux obtained from the best-fitting Gaussian models to avoid random fluctuations),
while the range for the $\lambda$4960 is the one of $\lambda$5008 divided by $1+\eta$.
The pixels with masks were excluded.
When $\eta$ changes, the wavelength range of the $\lambda$4960 line changes accordingly, and the degree of freedom may change when the $\lambda$4960 line contains masked pixels.
If so, we slightly adjusted the wavelength range to keep the degree of freedom constant.
An example of measuring $\eta$ and $A$ using the PM method is shown in the bottom row of Figure 6.

\begin{figure*}
\centering{
  \includegraphics[scale=0.8]{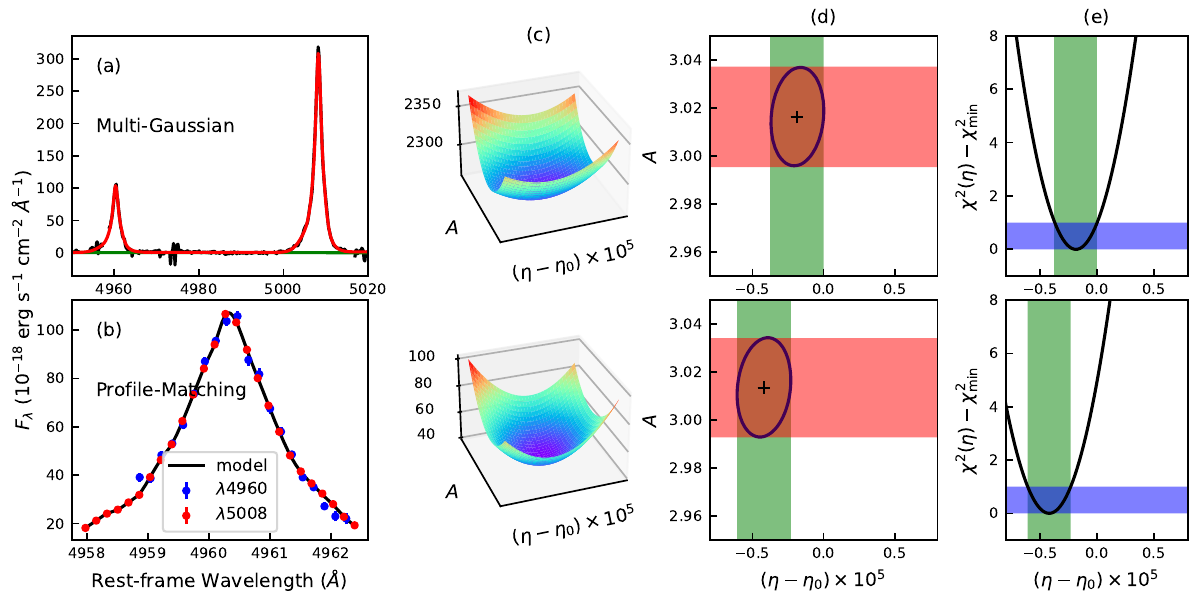}
  \caption{
  Examples of measuring $\eta$ and $A$ using the two methods.
  The MG method is shown in the upper row, and the PM method is shown in the bottom row.
  {\bf (a)}: Observed spectrum (black) and best-fitting multiple-Gaussian model (red).
  {\bf (b)}: The profile of $\lambda$4960 line (blue point), and that of $\lambda$5008 line which has been moved leftward and downsized (red point), and the best-fitting cubic spline function (black line).
  {\bf (c)}: The $\chi^2(\eta, A)$ surfaces obtained by the two methods.
  {\bf (d)}: The contours that yielding $\chi^2=\chi^2_{\rm min}+1$, and the resultant 1$\sigma$ confidence intervals of $\eta$ (green shade) and $A$ (red shade).
  {\bf (e)}: The minimum $\chi^2$ value for different $\eta$ (black line) and the 1$\sigma$ confidence interval of $\eta$ (green shade) corresponding to $\chi^2<\chi^2_{\rm min}+1$ (blue shade).
  }}
  \label{fig6}
\end{figure*}

For all the 86 spectra, the $\chi^2(\eta,A)$ surfaces obtained by the MG method look good: they are smooth, and their contours are close to ellipses in the vicinity of the minimum $\chi^2$, as shown in Figure 6(c) and 6(d).
This holds in the range of $\chi^2 < \chi_{\rm min}^2 + 1$ for all spectra and in the range of $\chi^2 < \chi_{\rm min}^2 + 10$ for most spectra.
Therefore, the measurements on $\eta$, $A$ and the corresponding statistical error are reliable.
However, only a small fraction of $\chi^2$ surfaces obtained by the PM method look good, and we showed two bad examples in Appendix B.
The PM method works well only for spectra with narrow \oiii doublet lines that have high SNR and almost bear no bad pixels.
We visually checked the $\chi^2$ surfaces and selected 16 LAE spectra and 1 QSO spectrum with which $\eta$ can be reliably measured.

The consistency between the two methods was checked by the measurements of $\eta$ from the 17 spectra. The $\eta$ values are showed in Figure 7(a), and their statistical errors $\sigma_{\rm stat}(\eta)$ in Figure 7(b).
The values of $\sigma_{\rm stat}(\eta)$ obtained from the two methods differ by less than 12\%.
To check whether the $\eta$ values were consistent, we calculated deviation $\Delta$ as:
\begin{equation}
\Delta=\frac{\eta({\rm MG})-\eta({\rm PM})}{\sqrt{\sigma_{\rm stat}^2(\eta,{\rm MG})+\sigma_{\rm stat}^2(\eta,{\rm PM})}}.
\end{equation}
The distribution of $\Delta$ is shown in Figure 7(c).
The average value is close to 0, and their absolute values are no more than 1.12.
These indicate that the results from the two methods agree with each other.
As the MG method works well for all the spectra in our sample, we adopted the measurements using it as the final result.
The best-fitting Multiple-Gaussian models are displayed in Appendix C.
The number of Gaussian functions, the best estimates and the statistical errors of $\eta$ and $A$ are listed in Table 3, 4 and 5.
As the measurements of the six spectra of SA enormously contributed to the final results, we also list the estimates of $\eta$ and $A$ using the PM method in Table 3 for comparison.

\begin{figure}
\centering{
  \includegraphics[scale=0.8]{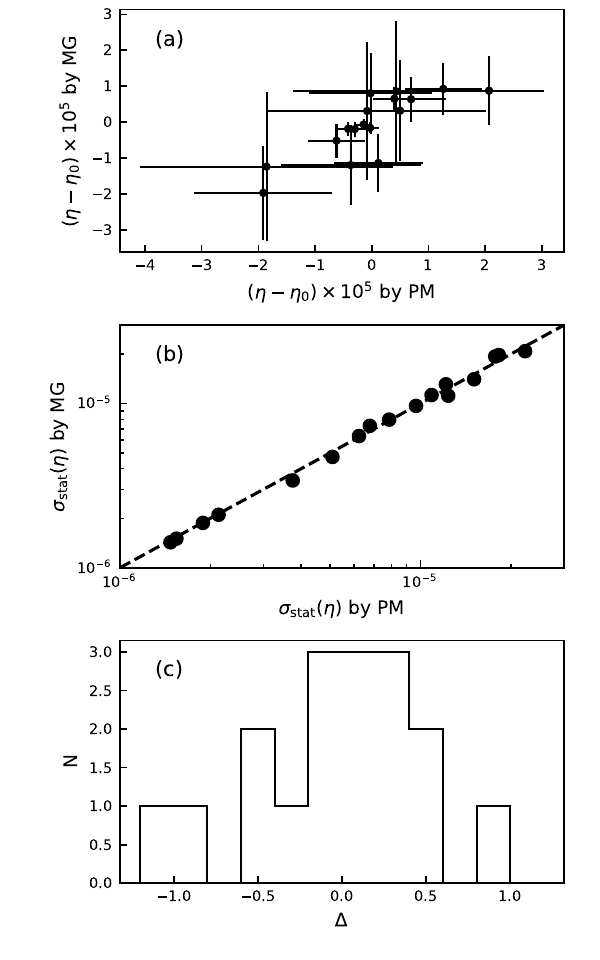}
  \caption{
  Comparison of the $\eta$ values measured using the two methods for the 17 spectra.
  {\bf (a)}: Comparison of $\eta$ values measured using the two methods.
  {\bf (b)}: Comparison of the statistical errors.
  {\bf (c)}: Distribution of $\Delta$ parameter, which characterizes the degree of deviation.
  }}
  \label{fig7}
\end{figure}

\subsection{Uncertainties from the Wavelength Calibration}

The wavelength calibration for the XShooter NIR Arm has a systematic error of $\Delta\lambda=0.04$\AA\ \citep[recorded in the Xshooter pipeline,][]{Modigliani2010}.
As $\eta \equiv (\lambda_2/\lambda_1-1)$, assuming that the wavelengths of $\lambda$4960 and $\lambda$5008 lines in the observer's frame both have a systematic error of $\Delta\lambda=0.04$ \AA, we obtained:
\begin{equation}
\sigma_{\rm sys}(\eta) = \frac{\sqrt{\lambda_1^2+\lambda_2^2}}{\lambda_1^2} \Delta\lambda
   = \frac{1.146\times10^{-5}}{1+z}.
\end{equation}
The total error $\sigma(\eta)$ can be calculated as:
\begin{equation}
\sigma(\eta) = \sqrt{ \sigma_{\rm stat}^2(\eta) + \sigma_{\rm sys}^2(\eta)}
\end{equation}
The added systematic errors will be on the order of $10^{-6}\sim10^{-5}$.
We checked whether or not these systematic errors should be considered using 17 spectra yielding $\sigma_{\rm stat}(\eta)<10^{-5}$.
If $\eta$ values from the 17 spectra were not to suffer from the systematic errors and their current statistical errors were enough to be responsible for their scatter around $\eta_0$, the values of $(\eta-\eta_0)/\sigma_{\rm stat}(\eta)$ would obey the standard normal distribution with the standard deviation around 1.
However, using the actual measured $\eta$, we obtained the standard deviation of $(\eta-\eta_0)/\sigma_{\rm stat}(\eta)$ to be 1.28, indicating that the probability of the current statistical errors being account for the scatter of $\eta$ around $\eta_0$ to be only 4.7\%.
Therefore, the systematic errors caused by the uncertainties of raising from the wavelength calibration should be considered.
After considering the systematic errors, the standard deviation becomes 1.04, consistent with the prediction of a standard normal distribution.
Therefore, involving the above systematic errors is reasonable, and we list the systematic errors of $\eta$ for all spectra in Tables 3, 4 and 5.

\begin{table*}
\caption{Measurements of $\eta$ and $A$ for the 6 spectra of SA (z=2.37).}
\small
\begin{tabular}{cccccccc}
\hline
ObjID & $N_G$ & \multicolumn{2}{c}{$(\eta-\eta_0)\times10^6$} & \multicolumn{2}{c}{$A$} & \multicolumn{2}{c}{$\Delta\alpha/\alpha\times10^4$} \\
      &       & MG & PM & MG &PM & MG & PM \\
\hline
SA1 & 4 & $-1.5\pm1.5\pm3.4$ & $-0.3 \pm1.5 \pm3.4$ & $2.99\pm0.02$ & $3.00\pm0.02$ & $-0.8\pm1.9 $ & $-0.2\pm1.9 $ \\
SA2 & 3 & $-0.7\pm1.4\pm3.4$ & $-1.4 \pm1.5 \pm3.4$ & $3.07\pm0.02$ & $3.07\pm0.02$ & $-0.4\pm1.9 $ & $-0.7\pm1.9 $ \\
SA3 & 4 & $-1.8\pm1.8\pm3.4$ & $-4.2 \pm1.9 \pm3.4$ & $3.02\pm0.03$ & $3.01\pm0.02$ & $-0.9\pm2.0 $ & $-2.2\pm2.0 $ \\
SA4 & 3 & $6.4 \pm3.3\pm3.4$ & $4.0  \pm3.8 \pm3.4$ & $3.02\pm0.03$ & $3.05\pm0.03$ & $3.3 \pm2.5 $ & $2.0 \pm2.6 $ \\
SA5 & 4 & $-1.9\pm2.1\pm3.4$ & $-3.0 \pm2.1 \pm3.4$ & $2.99\pm0.03$ & $2.99\pm0.02$ & $-1.0\pm2.0 $ & $-1.5\pm2.1 $ \\
SA6 & 3 & $-5.1\pm4.6\pm3.4$ & $-6.2 \pm5.1 \pm3.4$ & $3.03\pm0.05$ & $3.03\pm0.04$ & $-2.6\pm3.0 $ & $-3.2\pm3.2 $ \\
\hline
\end{tabular}
\label{tab3}
\end{table*}

\begin{table*}
\caption{Measurements of $\eta$ and $A$ for 34 other LAE spectra}
\small
\begin{tabular}{ccccccc}
\hline
ProgID & ObjID & z & $N_G$ & $(\eta-\eta_0)\times10^5$ & $A$ & $\Delta\alpha/\alpha\times10^4$\\
\hline
 084.A-0303(C)  & J1000+0201  & 2.246 & 1 & $1.5  \pm1.7 \pm0.4$ & $3.27\pm0.55$ & $7   \pm9  $ \\
 088.A-0672(A)  & J0332-2759  & 2.205 & 1 & $-1.2 \pm2.3 \pm0.4$ & $2.99\pm0.36$ & $-5  \pm11 $ \\
                & J0332-2800  & 2.172 & 3 & $-2.3 \pm1.0 \pm0.4$ & $3.06\pm0.12$ & $-11 \pm5  $ \\
 091.A-0413(A)  & J2346+1247  & 2.327 & 2 & $0.1  \pm3.7 \pm0.3$ & $3.06\pm0.42$ & $0   \pm19 $ \\
                & J2346+1249  & 2.173 & 2 & $1.6  \pm1.4 \pm0.4$ & $3.28\pm0.16$ & $8   \pm7  $ \\
 093.A-0882(A)  & J0217-0511  & 2.243 & 1 & $0.4  \pm1.2 \pm0.4$ & $3.01\pm0.18$ & $1   \pm6  $ \\
                & J0217-0512  & 2.183 & 2 & $-1.9 \pm1.3 \pm0.4$ & $3.35\pm0.20$ & $-9  \pm6  $ \\
                & J0332-2745  & 2.311 & 3 & $0.9  \pm0.7 \pm0.3$ & $3.09\pm0.09$ & $4   \pm4  $ \\
                & J0332-2746  & 2.226 & 2 & $-0.5 \pm1.8 \pm0.4$ & $2.92\pm0.18$ & $-2  \pm9  $ \\
                & J2358-1014  & 2.261 & 2 & $0.2  \pm0.8 \pm0.4$ & $2.92\pm0.12$ & $0   \pm4  $ \\
 096.A-0348(A)  & J0217-0514  & 2.185 & 1 & $-2.1 \pm2.4 \pm0.4$ & $3.40\pm0.35$ & $-10 \pm12 $ \\
                & J0217-0515  & 2.065 & 2 & $-0.5 \pm2.5 \pm0.4$ & $2.88\pm0.35$ & $-2  \pm13 $ \\
                & J0332-2741  & 2.312 & 2 & $0.8  \pm1.9 \pm0.3$ & $2.87\pm0.19$ & $4   \pm10 $ \\
 097.A-0153(A)  & J0145-0946  & 2.357 & 3 & $1.4  \pm0.8 \pm0.3$ & $3.06\pm0.08$ & $7   \pm4  $ \\
                & J0209-0005A & 2.167 & 1 & $0.1  \pm1.4 \pm0.4$ & $3.13\pm0.17$ & $0   \pm7  $ \\
                & J0209-0005B & 2.187 & 2 & $2.5  \pm2.6 \pm0.4$ & $3.24\pm0.23$ & $12  \pm13 $ \\
 099.A-0254(A)  & J0217-0502  & 2.209 & 1 & $1.8  \pm2.8 \pm0.4$ & $2.73\pm0.45$ & $9   \pm14 $ \\
 099.A-0758(A)  & J0332-2746  & 2.171 & 2 & $0.8  \pm1.1 \pm0.4$ & $3.10\pm0.21$ & $4   \pm5  $ \\
                & J0332-2749  & 2.033 & 2 & $2.1  \pm1.8 \pm0.4$ & $2.82\pm0.16$ & $10  \pm9  $ \\
 0101.B-0262(A) & J0014-3024  & 2.016 & 1 & $0.3  \pm1.9 \pm0.4$ & $3.51\pm0.48$ & $1   \pm9  $ \\
                & J0120-5143  & 1.295 & 2 & $3.6  \pm2.0 \pm0.5$ & $3.34\pm0.42$ & $18  \pm10 $ \\
                & J0131-1335  & 2.467 & 1 & $-0.2 \pm1.7 \pm0.3$ & $3.12\pm0.21$ & $-1  \pm8  $ \\
                & J2129+0006  & 2.405 & 1 & $0.5  \pm2.4 \pm0.3$ & $2.59\pm0.20$ & $2   \pm12 $ \\
                & J2129-0741  & 2.274 & 1 & $-1.9 \pm1.8 \pm0.3$ & $3.09\pm0.19$ & $-10 \pm9  $ \\
                & J2140-2339  & 2.486 & 2 & $0.9  \pm0.9 \pm0.3$ & $3.14\pm0.14$ & $4   \pm5  $ \\
                & J2248-4431  & 2.065 & 2 & $-1.2 \pm2.0 \pm0.4$ & $3.30\pm0.34$ & $-6  \pm10 $ \\
 0101.B-0779(A) & J1000+0236  & 3.423 & 1 & $0.6  \pm1.7 \pm0.3$ & $3.17\pm0.21$ & $3   \pm8  $ \\
 0102.A-0391(A) & J0332-2745  & 3.211 & 2 & $-0.0 \pm1.4 \pm0.3$ & $3.37\pm0.21$ & $0   \pm7  $ \\
 0102.A-0652(A) & J0217-0502  & 2.209 & 2 & $-1.2 \pm1.1 \pm0.4$ & $3.18\pm0.28$ & $-6  \pm5  $ \\
 0103.B-0446(A) & J0224-0449  & 2.550 & 2 & $-1.1 \pm0.8 \pm0.3$ & $3.04\pm0.13$ & $-5  \pm4  $ \\
                & J0332-2752  & 2.566 & 1 & $-1.5 \pm3.2 \pm0.3$ & $2.83\pm0.21$ & $-7  \pm16 $ \\
                & J1001+0206  & 2.435 & 2 & $8.1  \pm9.5 \pm0.3$ & $2.95\pm0.41$ & $41  \pm48 $ \\
 0104.A-0236(A) & J0332-2746  & 2.171 & 1 & $0.3  \pm1.4 \pm0.4$ & $2.86\pm0.19$ & $1   \pm7  $ \\
 0104.A-0236(B) & J0332-2746  & 2.171 & 1 & $-1.1 \pm1.8 \pm0.4$ & $3.35\pm0.25$ & $-5  \pm9  $ \\
\hline
\end{tabular}
\label{tab4}
\end{table*}

\begin{table*}
\caption{Measurements of $\eta$ and $A$ for the 46 QSO spectra}
\small
\begin{tabular}{ccccccc}
\hline
ProgID & ObjID & z & $N_G$ & $(\eta-\eta_0)\times10^5$ & $A$ & $\Delta\alpha/\alpha\times10^4$\\
\hline
 086.B-0320(A)  & J1242+0232  & 2.221 & 3 & $-6.1 \pm7.1 \pm0.4$ & $3.36\pm0.34$ & $-31 \pm36 $ \\
 087.A-0610(A)  & J0914+0109  & 2.139 & 1 & $6.0  \pm9.4 \pm0.4$ & $3.46\pm0.25$ & $30  \pm48 $ \\
 087.B-0229(A)  & J1512+1119  & 2.100 & 4 & $-0.5 \pm0.4 \pm0.4$ & $2.96\pm0.03$ & $-2  \pm2  $ \\
 088.B-1034(A)  & J0148+0003  & 1.481 & 2 & $1.0  \pm10.0\pm0.5$ & $3.66\pm0.54$ & $4   \pm51 $ \\
                & J0240-0758  & 1.533 & 2 & $2.8  \pm4.6 \pm0.5$ & $3.13\pm0.23$ & $14  \pm23 $ \\
                & J0323-0029  & 1.625 & 3 & $-1.0 \pm3.0 \pm0.4$ & $3.45\pm0.33$ & $-5  \pm15 $ \\
                & J0842+0151  & 1.496 & 2 & $-7.2 \pm6.8 \pm0.5$ & $3.90\pm0.30$ & $-36 \pm35 $ \\
                & J1005+0245  & 1.484 & 4 & $5.1  \pm4.8 \pm0.5$ & $3.14\pm0.26$ & $26  \pm24 $ \\
 089.B-0275(A)  & J1226-0006  & 1.126 & 2 & $-4.4 \pm4.1 \pm0.5$ & $3.11\pm0.16$ & $-22 \pm21 $ \\
 089.B-0936(A)  & J0904-2552  & 1.644 & 2 & $-1.8 \pm5.1 \pm0.4$ & $3.61\pm0.34$ & $-9  \pm26 $ \\
                & J1107-4449  & 1.598 & 2 & $-7.4 \pm7.2 \pm0.4$ & $3.42\pm0.26$ & $-38 \pm36 $ \\
 089.B-0951(A)  & J0958+0251  & 1.408 & 2 & $0.4  \pm3.2 \pm0.5$ & $3.22\pm0.17$ & $2   \pm16 $ \\
                & J1313-2716  & 2.198 & 3 & $-5.3 \pm4.6 \pm0.4$ & $3.36\pm0.18$ & $-27 \pm23 $ \\
 090.A-0830(A)  & J1002+0137  & 1.591 & 1 & $28.7 \pm12.8\pm0.4$ & $2.61\pm0.23$ & $147 \pm65 $ \\
                & J1003+0209  & 1.472 & 1 & $25.3 \pm19.9\pm0.5$ & $3.41\pm0.33$ & $130 \pm102$ \\
 090.B-0424(B)  & J0811+1720  & 2.332 & 4 & $1.9  \pm1.6 \pm0.3$ & $3.26\pm0.05$ & $9   \pm8  $ \\
 091.B-0900(A)  & J1313-2716  & 2.200 & 4 & $2.6  \pm3.8 \pm0.4$ & $3.21\pm0.20$ & $13  \pm19 $ \\
                & J1404-0130  & 2.515 & 2 & $1.0  \pm1.9 \pm0.3$ & $3.41\pm0.08$ & $5   \pm10 $ \\
                & J1544+0407  & 2.182 & 2 & $-3.9 \pm2.4 \pm0.4$ & $3.46\pm0.07$ & $-19 \pm12 $ \\
 092.A-0391(A)  & J0209-0438  & 1.132 & 4 & $0.6  \pm0.6 \pm0.5$ & $3.14\pm0.05$ & $3   \pm4  $ \\
 092.B-0860(A)  & J1154-0215  & 2.181 & 2 & $2.2  \pm3.3 \pm0.4$ & $3.37\pm0.16$ & $11  \pm16 $ \\
 093.B-0553(A)  & J0940+0034  & 2.332 & 2 & $8.1  \pm6.6 \pm0.3$ & $3.36\pm0.28$ & $41  \pm34 $ \\
                & J1019+0345  & 2.212 & 1 & $-20.0\pm10.0\pm0.4$ & $2.91\pm0.24$ & $-102\pm51 $ \\
                & J1323+0022  & 2.258 & 3 & $1.6  \pm2.5 \pm0.4$ & $3.19\pm0.11$ & $8   \pm13 $ \\
                & J2330-0216  & 2.496 & 2 & $-11.8\pm6.1 \pm0.3$ & $2.67\pm0.21$ & $-60 \pm31 $ \\
 094.B-0111(A)  & J0108+0134  & 2.099 & 2 & $-2.1 \pm3.3 \pm0.4$ & $3.36\pm0.13$ & $-10 \pm17 $ \\
 095.B-0507(A)  & J0114-0812  & 2.104 & 2 & $5.5  \pm6.0 \pm0.4$ & $2.23\pm0.12$ & $28  \pm30 $ \\
                & J2136-1631  & 1.659 & 4 & $0.2  \pm1.4 \pm0.4$ & $3.00\pm0.15$ & $0   \pm7  $ \\
 098.B-0556(A)  & J0111-3502  & 2.405 & 3 & $-0.5 \pm0.3 \pm0.3$ & $3.07\pm0.02$ & $-2  \pm2  $ \\
                & J0331-3824  & 2.435 & 2 & $0.5  \pm1.4 \pm0.3$ & $3.24\pm0.04$ & $2   \pm7  $ \\
 099.A-0018(A)  & J1338+0010  & 2.300 & 2 & $-10.9\pm6.7 \pm0.3$ & $3.09\pm0.23$ & $-55 \pm34 $ \\
 099.B-0118(A)  & J0020-1540  & 2.020 & 3 & $-1.8 \pm2.5 \pm0.4$ & $2.87\pm0.08$ & $-9  \pm13 $ \\
 0101.A-0528(A) & J2124-4948  & 2.480 & 2 & $4.2  \pm2.4 \pm0.3$ & $3.28\pm0.21$ & $21  \pm12 $ \\
 0101.A-0528(B) & J0607-6031  & 1.097 & 3 & $8.5  \pm3.8 \pm0.5$ & $3.15\pm0.17$ & $43  \pm19 $ \\
 0101.B-0739(A) & J1352+1302  & 1.600 & 1 & $8.8  \pm12.2\pm0.4$ & $4.11\pm0.34$ & $45  \pm62 $ \\
                & J1616+0931  & 1.468 & 2 & $0.8  \pm3.5 \pm0.5$ & $3.24\pm0.13$ & $4   \pm18 $ \\
                & J2239+1222  & 1.484 & 3 & $5.8  \pm6.7 \pm0.5$ & $3.41\pm0.22$ & $29  \pm34 $ \\
 0102.A-0335(A) & J0259+1635  & 2.160 & 3 & $-0.2 \pm0.2 \pm0.4$ & $3.09\pm0.01$ & $-1  \pm2  $ \\
                & J0630-1201  & 3.344 & 2 & $9.9  \pm9.7 \pm0.3$ & $3.48\pm0.34$ & $50  \pm49 $ \\
                & J1042+1641  & 2.519 & 4 & $-0.4 \pm0.5 \pm0.3$ & $2.84\pm0.01$ & $-1  \pm2  $ \\
 0103.A-0253(A) & J2245-2945  & 2.379 & 2 & $6.9  \pm3.7 \pm0.3$ & $2.96\pm0.39$ & $35  \pm18 $ \\
 189.A-0424(A)  & J1018+0548  & 3.512 & 2 & $10.8 \pm6.9 \pm0.3$ & $3.21\pm0.20$ & $55  \pm35 $ \\
                & J1042+1957  & 3.628 & 1 & $30.8 \pm16.1\pm0.2$ & $2.81\pm0.24$ & $158 \pm82 $ \\
                & J1117+1311  & 3.623 & 1 & $-4.3 \pm11.8\pm0.2$ & $2.87\pm0.25$ & $-22 \pm60 $ \\
                & J1312+0841  & 3.735 & 1 & $15.4 \pm10.9\pm0.2$ & $2.67\pm0.22$ & $79  \pm56 $ \\
                & J1621-0042  & 3.703 & 1 & $-6.9 \pm9.2 \pm0.2$ & $3.77\pm0.24$ & $-35 \pm47 $ \\
\hline
\end{tabular}
\label{tab5}
\end{table*}

\subsection{Robustness of The Measurements}

Results from the final sample were checked statistically.
We first checked if $\eta$ correlated with $A$ or not.
The distribution of $\eta$ and $A$ measured from the 86 spectra is shown in Figure 8.
The Pearson correlation coefficients of the LAE subsample, the QSO subsample and the whole sample are $-0.13$, $-0.15$ and $-0.11$, respectively, indicating no correlation.
This is consistent with theoretical expectations, as \cite{Bahcall2004} discussed.
Using a bootstrap method described in Section 4.1, we calculated the average values of $A$ of the LAE and the QSO subsamples, which are $3.03\pm0.03$ and $3.05\pm0.06$, respectively.
These values are consistent with $2.99\pm0.02$ measured by \cite{Bahcall2004} and $2.96\pm0.02$ measured by \cite{Albareti2015} within 3$\sigma$, and are also consistent with the theoretical value of 2.98 \citep{Storey2000}.

\begin{figure}
\centering{
  \includegraphics[scale=0.8]{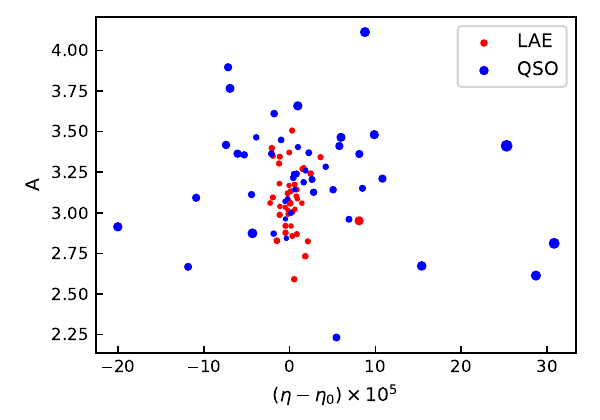}
  \caption{
  Plots of $\eta$ and $A$ of the final sample measured by the MG method.
  LAE results are in red, and QSO results are in blue.
  The size of the dots represents the error of $\eta$ (larger for greater error).
  We did not show the error of $A$, but those with greater error of $\eta$ generally have greater error of $A$.
  }}
  \label{fig8}
\end{figure}

We then checked whether the measured $\sigma_{\rm stat}(\eta)$ is related to the SNR and FWHM of the \oiii lines as expected.
As can be seen in Figure 9, $\sigma_{\rm stat}(\eta)$ is inversely related to SNR$_{4960}$ for similar FWHM and is positively related to FWHM for similar SNR$_{4960}$.
We verified this finding through numerical simulations, as detailed in Appendix D.
Numerical simulations give quantitative correlations, as expressed in formulas D1 and D2.
In brief, $\sigma_{\rm stat}(\eta)$ is roughly inversely proportional to SNR$_{4960}$ and positively proportional to FWHM.
In Figure 9, we show the relation with FWHMs of 130 and 610 km s$^{-1}$, which are the median FWHM values of LAE and QSO subsamples, respectively.
For both the subsamples, the measurements are in good agreement with the theoretical expectations with scatters less than 0.4 dex.

\begin{figure}
\centering{
  \includegraphics[scale=0.8]{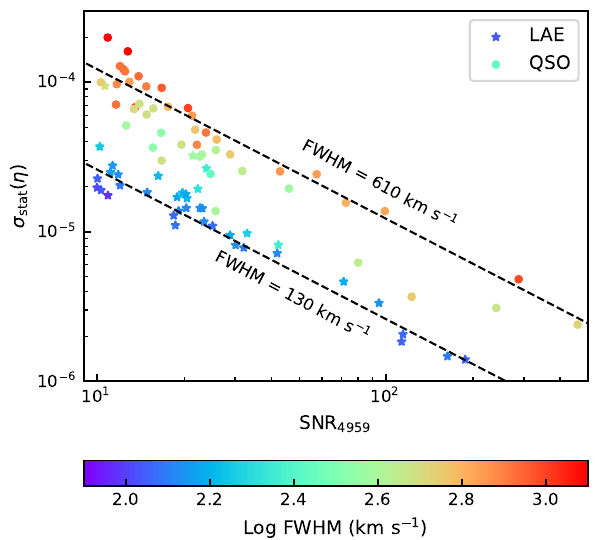}
  \caption{
  Relation between $\sigma_{\rm stat}(\eta)$ and SNR$_{4960}$.
  The pentagrams represent LAEs, and the dots represent QSOs.
  The colour represents the \oiii FWHM (see the colour bar below).
  The two black dashed lines show theoretical expectations for FWHMs of 130 and 610 km s$^{-1}$, respectively.
  }}
  \label{fig9}
\end{figure}

We finally checked whether $\chi=(\eta-\eta_0)/\sigma(\eta)$ obeys the standard normal distribution or not, which should be the case if the true value of all $\eta$ were $\eta_0$ and $\sigma(\eta)$ could be account for the uncertainty in the measurement of $\eta$.
Figure 10 shows the distribution of $\chi$.
The mean value and standard deviation of $\chi$ and the mean value of $\chi^2$ are 0.11, 1.02 and 1.04, respectively.
These values are consistent with the theoretical values with a sample size 86, which are $0\pm0.11$, $1\pm0.08$ and $1\pm0.15$ (68.3\% confidence level), respectively.
We made a Kolmogorov-Smirnov (KS) test.
We calculated a KS statistic value of 0.11, corresponding to a probability of 0.21, $>$0.05.
Thus, we concluded that the distribution of $\chi$ obeys the standard normal distribution.

\begin{figure}
\centering{
  \includegraphics[scale=0.8]{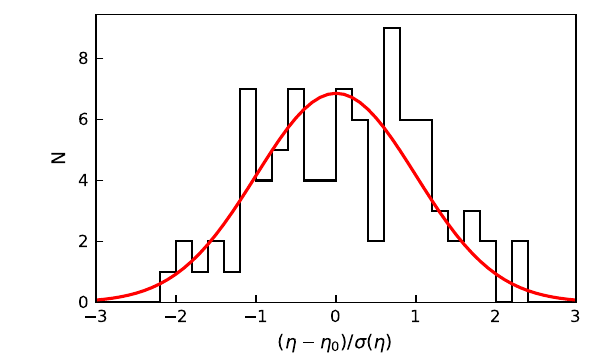}
  \caption{
  The distribution of $\chi$ and a comparison with the standard normal distribution (red line).
  }}
  \label{fig10}
\end{figure}

\section{Results of Variation in the Fine Structure Constant}

We converted all the measurements of $\eta$ and their errors $\sigma(\eta)$ into values and errors of $\Delta\alpha/\alpha$ using equation (8).
The results are listed in Tables 3, 4, and 5, and shown in Figure 11.
None of the measurements with all 86 spectra shows a deviation of $\Delta\alpha/\alpha$ from 0 by more than 3$\sigma$, and accuracies are between $2\times10^{-4}$ and $10^{-2}$.

According to equation (8), $\Delta\alpha/\alpha$ is proportional to $\eta-\eta_0$.
Hence $\chi=(\eta-\eta_0)/\sigma(\eta)$ defined in section 3.3 equals $\frac{\Delta\alpha/\alpha}{\sigma(\Delta\alpha/\alpha)}$, and also represents the deviation of $\Delta\alpha/\alpha$ from 0.
In section 3.3, we have shown that $\chi$ for the 86 spectra obeys the standard normal distribution.
This is consistent with the assumption that the truth value of $\Delta\alpha/\alpha$ is 0, and the errors were estimated reasonably.

\begin{figure}
\centering{
  \includegraphics[scale=0.8]{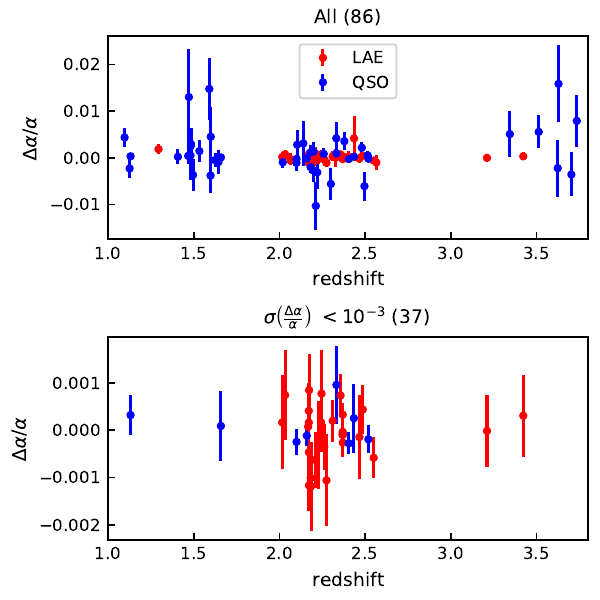}
  \caption{
  The upper panel shows the measurements of $\Delta\alpha/\alpha$ using all 86 spectra.
  The lower panel shows those using 37 spectra with errors $<10^{-3}$.
  }}
  \label{fig11}
\end{figure}

\subsection{Averages}

We calculated average values for the final sample and different subsamples using two methods.
The first is the weighted average method (WM).
For values $x_i=(\Delta\alpha/\alpha)_i$ and errors $\sigma_i$, we calculated the weight as:
\begin{equation}
w_i = \frac{1}{\sigma_i^2}.
\end{equation}
Thus, the weighted average is:
\begin{equation}
\bar{x}_{\rm WM} = \frac{\sum_i w_i x_i}{\sum_i w_i},
\end{equation}
and its error is:
\begin{equation}
\sigma(\bar{x})_{\rm WM} = \frac{ \sqrt{\sum_i (w_i \sigma_i)^2} }{\sum_i w_i}.
\end{equation}
The second is the bootstrap method \citep[BS, e.g.,][]{Bahcall2004}.
For a sample containing $N$ elements, we randomly generated $10^5$ fake bootstrap samples, each containing $N$ elements by put-back sampling.
We calculated the weighted averages of these bootstrap samples with equations (12) to (14).
These $10^5$ weighted averages typically obey a Gaussian distribution.
We adopted the mean value and standard deviation of this Gaussian distribution $\bar{x}_{\rm BS}$ and $\sigma(\bar{x})_{\rm BS}$ as the bootstrap average and its error.
Both WM and BS methods have their advantages and disadvantages.
On the one hand, when the errors $\sigma_i$ are estimated inaccurately, the error of average value by the BS method is more accurate.
For example, suppose all the errors are overestimated or underestimated by a constant factor; the error by the BS method will not change, but the error by the WM method will be overestimated or underestimated correspondingly.
On the other hand, when the sample size $N$ is small or the errors $\sigma_i$ vary significantly from one another, the weighted averages of bootstrap samples in the BS approach may not obey Gaussian distribution.
In this case, the results of the WM method are more reliable.
For each sample or subsample, we calculated the average value of $\Delta\alpha/\alpha$ and its error using both methods, and the results are listed in Table 6.

The average of $\Delta\alpha/\alpha$ of all the 86 spectra is $(-3\pm6)\times10^{-5}$ and $(-3\pm5)\times10^{-5}$ using WM and BS methods, respectively.
The two values are consistent and indicate no deviation between $\Delta\alpha/\alpha$ and 0 in the $6\times10^{-5}$ error level.
We calculated the averages of the LAE and QSO sub-samples and further divided the former into SA and other LAE sub-samples and calculated the averages.
For each subsample, the averages by both two methods agree with 0.

We found that for the SA subsample, the error of average by the BS and WM methods have a great difference, while for other subsamples, the difference is slight.
This may be because the SA subsample has a small size $N$.
Because we have shown that the errors $\sigma(\Delta\alpha/\alpha)$ are estimated reasonably, we accept the results by the WM method as the final results.

The above results are based on $\eta$ measured by the MG method.
As described in section 3.1, there are 17 spectra with which $\eta$ can be measured by the PM method, including 6 spectra of SA.
The averages of $\Delta\alpha/\alpha$ measured by the PM method are shown in Table 6, which are also consistent with 0.
This indicates that the coincidence of $\Delta\alpha/\alpha$ and 0 does not depend on how we measured $\eta$.

\begin{table}
\centering{
\caption{Average $\Delta\alpha/\alpha$ for different samples}
\begin{tabular}{cccc}
\hline
subsample & N & \multicolumn{2}{c}{$\Delta\alpha/\alpha\times10^5$}\\
          &   & WM  & BS \\
\hline
All           & 86 & $-3.1\pm6.0$  & $-3.1\pm5.0$ \\
All QSO       & 46 & $-7.6\pm11.1$ & $-7.3\pm10.6$ \\
All LAE       & 40 & $-1.2\pm7.2$  & $-1.3\pm6.2$ \\
SA            & 6  & $-4.1\pm8.7$  & $-4.3\pm5.6$ \\
Other LAE     & 34 & $4.9\pm12.7$ & $5.1\pm13.4$ \\
\hline
SA PM         & 6  & $-9.1\pm8.9$  & $-9.2\pm5.3$ \\
All good PM   & 17 & $-2.6\pm7.9$  & $-2.6\pm6.8$ \\
\hline
$1.09<z<1.66$ & 18 & $48\pm31$ & $47\pm31$ \\
$2.01<z<2.57$ & 60 & $-5.7\pm6.2$  & $-5.8\pm5.0$ \\
$3.21<z<3.74$ & 8  & $39\pm55$ & $41\pm119$ \\
z2 without SA & 54 & $-7.3\pm8.8$ & $-7.3\pm8.3$ \\
\hline
\end{tabular}}
\label{tab6}
\end{table}

Due to telluric absorption bands, the whole sample can be naturally divided into three subsamples in three redshift ranges.
The averages of $\Delta\alpha/\alpha$ of the three subsamples, listed in Table 6, are all consistent with 0.
The errors of these averages are $3\times10^{-4}$, $6\times10^{-5}$, and $5\times10^{-4}$, respectively.
Because the results obtained from the spectra of SA ($z=2.37$) have high precision, one may worry that the average value in the redshift range of $2.01<z<2.57$ is greatly affected by SA and cannot be treated as the representative value of this redshift range.
Thus, we calculated the average of $\Delta\alpha/\alpha$ obtained from other 54 spectra in this redshift range, and the result agrees with 0 with an error of $9\times10^{-5}$.

\subsection{Variation over Time}

We do not find any variance of $\Delta\alpha/\alpha$ over time because it agrees with 0 in all the redshift ranges.
We limited the rate at which $\alpha$ may vary, which can be used to constrain cosmological models further.

The redshifts of LAEs and QSOs in our sample are in the range of 1.097--3.735, corresponding to look-back time ($t_{\rm LB}$) of 8.2--12.0 billion years, or the age of the universe of 1.7--5.5 billion years.
We assumed that $\alpha$ varies uniformly over time from then to now:
\begin{equation}
\frac{\Delta\alpha}{\alpha} = k t_{\rm LB}.
\end{equation}
The $k$ here means the same as $\alpha^{-1}{\rm d}\alpha/{\rm d}t$ in \cite{Bahcall2004}.
We fit the $\Delta\alpha/\alpha$ measurements of all the 86 spectra, and obtained $k=(-3\pm6)\times10^{-15}$ yr$^{-1}$.

We then included $\Delta\alpha/\alpha$ measured by \cite{Albareti2015} using SDSS QSO spectra (see their Table 3) into the rate limitation.
Note that we did not use the data in the redshift bin 0.580--0.625 in their results because it may be seriously affected by the SELs.
If only using measurements at $z<1$, the result is $k=(2\pm5)\times10^{-15}$ yr$^{-1}$.
And by combining the measurements at $z<1$ and our measurements at $1.0<z<3.8$, we obtained $k=(0\pm4)\times10^{-15}$ yr$^{-1}$.
This is by far the most precise limitation of the rate using the \oiii doublet method.
Moreover, our results extend the limitation to an era when the universe is only 2 to 5 billion years old.

\begin{figure}
\centering{
  \includegraphics[scale=0.8]{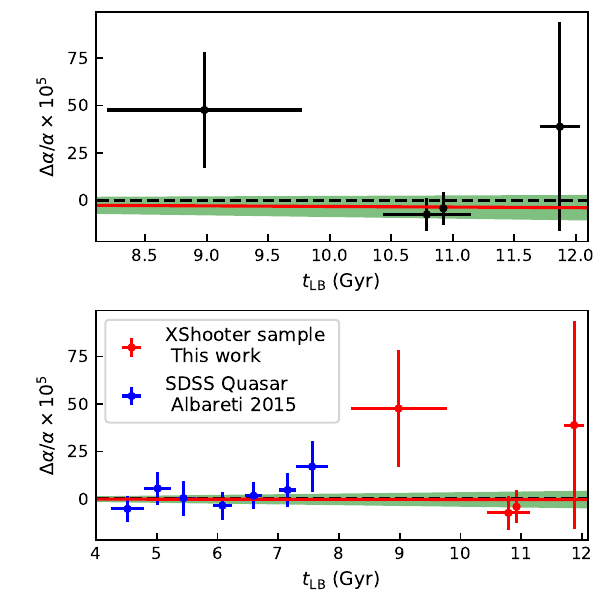}
  \caption{
  Tests of whether $\alpha$ changes over time.
  The upper panel shows the result using only our sample, and the bottom panel shows the result using both our sample and the measurements using SDSS QSO spectra by Albareti et al. (2015). 
  The best-fitting model assuming that $\alpha$ varies at a uniform rate over time is shown in red, and the 1$\sigma$ confidence interval is shown in green shade.
  }}
  \label{fig12}
\end{figure}

\section{Discussions}

\subsection{Considerations on Selecting LAE and QSO for $\alpha$ Measurement}

We discuss the reliability and necessity of our final sample selection criteria.

First, we only selected spectra where SNRs of both the \oiii doublet lines are greater than 10.
An SNR of $>$10 guarantees that the contours of the $\chi^2(\eta,A)$ surface near the minimum value are close to ellipses, and hence, the probability density distributions of $\eta$ and $A$ are close to Gaussian, which is convenient for subsequent analysis.
In addition, the $\eta$ measured from spectra with low \oiii SNR have large errors and hence have small weights when calculating average and rate.
Therefore, discarding them hardly affects the precision of the final results.
Some previous work studying $\alpha$ variation with optical QSO spectra used the SNR of $\lambda$5008 as a criterion for sample selection.
We suggest the SNR of $\lambda$4960 line to be included in the criteria when using NIR spectra, because it is not strictly correlated with the SNR of $\lambda$5008 line (Figure D1) due to the influence of telluric lines.

Second, we added masks to pixels affected by SELs, TALs, cosmic rays and bad CCD pixels, and excluded these pixels when measuring line wavelengths.
In addition, we excluded spectra where many pixels near \oiii doublet are affected, using a mask index $I_{\rm mask}$ for quantification.
Note that this approach differs from previous works using optical spectra, where pixels affected by SELs and TALs were included in the measurements but were given lower weights due to larger errors.
In analysis concerning NIR spectra, systematic errors caused by corrections for SELs and TALs must be addressed, and therefore, the errors of the pixels where strong SELs and TALs are located may be significantly underestimated.
Fortunately, the VLT/XShooter spectra have high spectral resolutions, so the impacting ranges of single SEL or TAL are narrow.
For a typical XShooter spectrum, the proportion of pixels strongly affected by SELs or TALs is generally between 10\% and 20\%, except for wavelength ranges where telluric absorption lines are dense.
Most spectra can pass the criterion that the sum of the $I_{\rm mask}$ of the doublet lines is less than 0.5.
Even if the affected pixels are directly masked for these spectra, the remaining pixels are sufficient to achieve reliable wavelength measurements.

Finally, we had two additional requirements for QSO spectra compared with previous works.
One is that there should be no strong Fe II bump, and the other is that the doublet lines should not blend severely.
For these two requirements, we measured Fe II index $I_{\rm FeII}$ and blending index $I_{\rm blend}$ as the criteria.
We tested whether the two criteria are reasonable by examining the spectra they excluded.
For a continuum-subtracted spectrum, we obtained the profiles of the doublet emission lines in wavelength ranges corresponding to a velocity range of $-600$ to $600$ km s$^{-1}$.
We made linear interpolation for the profile of $\lambda$5008 to align it with the profile of $\lambda$4960 in velocity space, and then calculated the Pearson correlation coefficient between them.
This coefficient represents the consistency of the shapes of the doublet lines.
In Figure 13, we show the coefficients of QSO spectra excluded by the Fe II or the blending criterion, those of QSO spectra in the final sample, and the $A$ values measured using these spectra.
We only show the results of spectra with SNR$_{5008}>50$ to ensure the accuracy of the parameters.
We found that the doublet profiles have a good correlation in the final sample as the coefficients have a median value of 0.95 and are all greater than 0.87, and the measurements of $A$ are all around the theoretical value of 3.
However, the correlation between the doublet profiles is not as good in QSO spectra excluded by the Fe II criterion, as the coefficients have a median value 0.83.
Also, a large $A$ value of $>$4 is measured for more than half of these excluded spectra.
The likely reason is that the Fe II bump under the \oiii doublet causes the continuum to be fit improperly.
The correlation can be poor in QSO spectra excluded by the blending criterion as more than half of coefficients are $<$0.5, and the measured values of $A$ spread over a wide range.
The likely reason is the mistaken decomposition of the very broad \oiii emission components.
The anomalies in the correlation coefficients and measurements of $A$ indicate that these spectra are unsuitable for the measurements of doublet wavelengths.
Hence, our criteria are necessary and practical.

\begin{figure}
\centering{
  \includegraphics[scale=0.8]{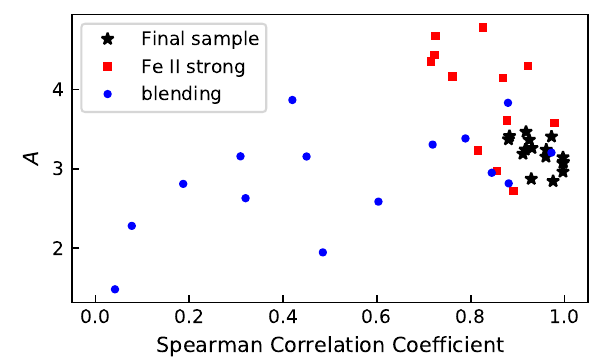}
  \caption{
  We present the correlation coefficients of the doublet line profiles and the measurements of $A$ for QSO spectra in the final sample (black pentagram), excluded by the Fe II criterion (red box) and excluded by the blending criterion (blue origin).
  }}
  \label{fig13}
\end{figure}

According to the initial redshift limit of $1.07<z<3.77$, 95 LAE spectra and 601 QSO spectra were selected.
However, the final sample only contains 40 LAE and 46 QSO spectra.
The passing rate of LAE and QSO subsamples are 42\% and 8\%, respectively.
About 20\% of the LAE spectra are excluded because of SELs: in some cases, the mask index is too large, and in other cases, the SELs result in insufficient SNR of \oiii.
This fraction is similar to the fraction of pixels strongly affected by SELs, which is 10\%--20\% as we previously estimated.

QSO spectra have a lower pass rate.
A few spectra are excluded due to observational factors, such as short exposure time causing insufficient \oiii SNR or inappropriate redshift causing \oiii to fall into the telluric absorption bands.
These factors are similar for QSOs and LAEs.
However, more than half of the spectra are excluded because of the QSOs' intrinsic properties, such as the weak \oiii relative to the continuum, the blending \oiii doublet, or a strong Fe II bump.
The composite of the 46 QSO spectra in the final sample, displayed in Figure 14, differs from that of general QSOs.
To demonstrate this difference, we also present other QSO composite spectra for comparison, including that of \cite{Selsing2016} using XShooter data of $1<z<2$ high-luminosity QSOs, and that of \cite{VandenBerk2001} using SDSS data (of $z<1$ QSOs for the wavelength range around \oiii).
The QSOs in our sample have much stronger \oiii lines, and the EW of \oiii lines in their composite spectrum can reach 31.9 \AA, much higher than those in the other two composite spectra (Table 7).
Also, they have weaker Fe II bumps as the Fe II indexes are smaller.
Thus, QSOs suitable for $\alpha$ measurement may belong to a particular type, accounting for a small proportion of the total number of QSOs.
This may be the main reason for the low pass rate of QSO spectra.
Assuming that the proportions of LAEs and QSOs excluded due to observational factors are similar, with a pass rate of $\sim40$\%, we estimated that only 20\% of the QSO spectra are suitable for $\alpha$ measurement.

\begin{figure}
\centering{
  \includegraphics[scale=0.8]{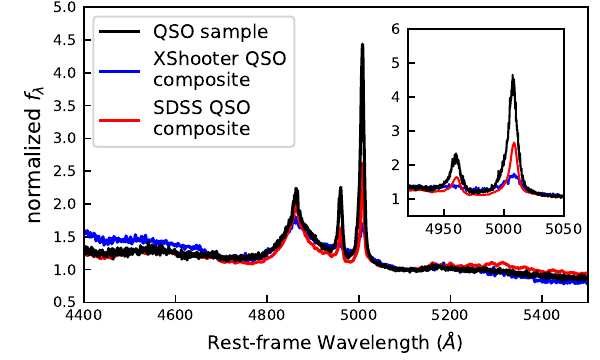}
  \caption{
  The composite spectrum of QSOs in our final sample (black).
  We show XShooter (blue) and SDSS (red) QSO composite spectra for comparison.
  All three spectra are normalized at 5100 \AA.
  The insert panel shows the partial enlargement near the \oiii doublet.
  }}
  \label{fig14}
\end{figure}

\begin{table}
\centering{
\caption{Comparison between our QSO sample and QSO composite}
\begin{tabular}{cccc}
\hline
          & our & XshQ & SDSSQ\\
\hline
z                  & 1.09--3.73  & 1--2.1 & $<1$ \\
$\nu L_\nu(5100\AA)$ (erg s$^{-1}$) & 8e45   & $\sim$1e47 & $\sim$3e44 \\
$\alpha_\lambda$ (4450--5500 \AA)   & $-$1.7 & $-$2.7     & $-$1.1 \\
$[$OIII] $\lambda$5007 EW ($\AA$)   & 31.9 & 10.0 & 13.4 \\
$[$OIII] FWHM (km s$^{-1}$) & 510  & 970  & 460 \\
$I_{\rm 4590}$ ($\AA$)      & 7.6  & 12.0 & 18.1 \\
$I_{\rm 5250}$ ($\AA$)      & 9.7  & 16.3 & 16.3 \\
\hline
\end{tabular}}
\label{tab7}
\end{table}

\subsection{Implications for Future Works}

We suggest that LAE spectra have more advantages than QSO spectra studying of $\alpha$ variation using the \oiii doublet method.
The reasons are as follows.

First, the \oiii doublet lines are narrower in LAE spectra than those in QSOs' spectra,
as the median values of the \oiii FWHM in the LAE and QSO subsamples are 130 and 610 km s$^{-1}$, respectively, with a difference of 4.7 times.
The value of the FWHM affects the statistical error of $\eta$, as narrated in Appendix D, where $\sigma_{\rm stat}(\eta) \propto {\rm FWHM^{3/2}}$ for the same flux.
Therefore, a difference of 4.7 times in the values of FWHM would lead to a 10 times difference in $\sigma_{\rm stat}(\eta)$ if other conditions were the same.
In our sample, the resultant precision of $\Delta\alpha/\alpha$ from the LAE and QSO subsamples are on the same level, for the higher luminosities of the \oiii lines in the QSOs' spectra compensate for their disadvantage in the FWHM.
The median \oiii luminosity of QSOs ($10^{43.7}$ erg s$^{-1}$) in our sample is 10 times that of LAEs ($10^{42.7}$ erg s$^{-1}$), as can be seen in Figure 4.

Second, the volume density of LAEs suitable for studying $\alpha$ variation is much higher than that of QSOs at $z>1$.
We calculated the volume density of LAEs with Ly$\alpha$ luminosity greater than the mean value of our LAEs, and that of QSOs with continuum luminosity greater than the mean value of our QSOs, both at $z\sim2$.
We extracted the spectra on the UVB arm of $z\sim2$ LAEs in our sample and measured the Ly$\alpha$ luminosities, which have a median value of $10^{43.1}$ erg s$^{-1}$.
Using the Ly$\alpha$ luminosity function of $z=2.2$ LAEs from \cite{Konno2016}, we estimated the volume density of LAEs with $L_{\rm Ly\alpha}>10^{43.1}$ erg s$^{-1}$ is $\sim5\times10^{-5}$ Mpc$^{-3}$.
We measured the G-band absolute magnitudes of the QSOs in our sample, with a median value of $-26.2$.
Using the continuum luminosity function of $2.2<z<2.6$ QSOs from \cite{Croom2009}, we estimated that the number density of QSOs with $g<-26.2$ is only $\sim5\times10^{-7}$ Mpc$^{-3}$.
In addition, only a small fraction of QSOs have strong and narrow \oiii lines and weak Fe II bumps.
We estimated this fraction at 20\% in section 5.1.
As a result, the number density of LAEs suitable for $\alpha$ measurement is 2--3 orders of magnitude higher than QSOs.

Finally, systematic errors may be larger and more challenging to analyze when measuring $\alpha$ with QSO spectra.
In this work, we considered two possible sources of systematic error for QSOs, the blending of \oiii doublet and Fe II bump.
We selected QSOs with \oiii not severely blended and with weak Fe II.
The resultant $\eta/\sigma(\eta)$ obeys the standard normal distribution, implying that the possible systematic errors from these two sources hardly affect the results of our work.
However, if the method is applied to future $\alpha$ measurements with higher precision, these systematic errors may become significant as other errors decrease.
In addition, theoretically, there are some other sources of systematic errors, such as the uncertainty of the shape of H$\beta$ broad emission lines, weak QSO emission lines, and others.
These errors may also affect future works using QSO spectra for accurate $\alpha$ measurement.
In LAE spectra, the continuum is weak and mainly comes from starlight.
Thus, the modelling of the LAE continuum is more reliable than the modelling of the QSO continuum.

We discuss the possible precision of future $\alpha$ measurements using spectroscopy of LAE and other starburst galaxies at $z<2$.
This work obtained an error of $\Delta\alpha/\alpha$ of $7\times10^{-5}$ using 40 XShooter spectra of LAEs.
The error is $9\times10^{-5}$ if only using 6 spectra of SA and is $1.3\times10^{-4}$ if only using the other 34 spectra.
To obtain an error of $\Delta\alpha/\alpha$ below $10^{-6}$ to test the measurements using the many-multiplet method, $N>50,000$ spectra similar to the SA's or $N>600,000$ spectra similar to other LAEs' are required, assuming that the error is inversely proportional to $\sqrt{N}$.
To achieve the required sample size, observing multiple spectra of starburst galaxies simultaneously using multi-object spectrometers is necessary.
The ongoing project Dark Energy Spectroscopic Instrument \citep[DESI,][]{DESI2016} may meet the sample size demand.
The DESI project is expected to observe about a dozen million emission line galaxies, of which $\sim$7 million at $z<0.95$, suitable for the study of $\alpha$ variation using \oiii doublet as the wavelength range is 3600--9800 \AA.
Spectra with sufficiently high \oiii SNRs may be on the order of $10^5\sim10^6$.
Because the sensitivity, resolution and sample size of DESI are all better than SDSS, the final accuracy of $\alpha$ measurement must be better than the existing results using SDSS spectra ($\sim2\times10^{-5}$), and might be comparable with the results using the many-multiplet method at present.

So far, the highest redshift at which $\alpha$ variation has been measured is 7.1 \citep{Wilczynska2020}.
Their work used the absorption lines in the VLT/XShooter spectra of QSO J1120+0641 at $z=7.085$.
Measuring $\alpha$ variation at higher redshift using QSO absorption lines is challenging because QSOs at $z>7$ are extremely rare.
Fortunately, a number of $z>7$ LAEs have been spectroscopically identified \citep[e.g.,][]{Vanzella2011,Stark2017,Jung2020,Endsley2021}.
Their spectral energy distributions suggest they have strong \oiii emission lines.
These LAEs can be used to constrain the variation of $\alpha$ at $z>7$ as long as deep mid-infrared spectroscopic observations are obtained because the \oiii wavelengths are redshifted to $>4$ $\mu$m.

The Near InfraRed Spectrograph (NIRSpec) mounted on James Webb Space Telescope \citep[JWST,][]{Gardner2006} can take spectrum in a wavelength range of 0.6--5.3 $\mu$m, and hence can observe \oiii emission lines of LAEs at $z<9.4$.
Depending on the observational mode, the spectral resolution $R$ can be 100, 1000 or 2700.
Assuming that the LAEs have an intrinsic \oiii FWHM of 130 km s$^{-1}$ and are observed with JWST/NIRSpec with $R$ of 1000 or 2700, we estimated the observed \oiii FWHM to be 300 or 170 km s$^{-1}$ using equation D3 in Appendix D.
Assuming an SNR of $\lambda$4960 line of 10, we further estimated that the precision of $\Delta\alpha/\alpha$ measurement could reach $(2\sim3)\times10^{-3}$.
Higher precision could be obtained if the SNR is higher or more than one LAE is observed.
This measurement will, for the first time, provide a constraint on the variation of $\alpha$ at $7.1<z<9.4$, corresponding to an age of the universe of only 500--700 million years.

\section{Conclusions}

We constructed a sample of 86 public NIR spectra observed by VLT/XShooter of LAE (40) and QSOs (46) at $1.09<z<3.73$.
The selection criteria of the sample concern the SNRs of the \oiii doublets and three indexes: the first describes to what extent the data is affected by SELs, TALs, cosmic rays and bad CCD pixels, the second describes the strength of Fe II bump in the QSO spectrum, and the third describes the blending of the \oiii doublet.
We measured the parameter $\eta$ describing the ratio of the wavelengths of the \oiii doublet.
We further inferred $\Delta\alpha/\alpha$, the difference between $\alpha$ in the distant universe and that in the laboratory.
When measuring $\eta$, we tried a Multiple-Gaussian method and a Profile-Matching method, in which the former applies to all spectra (86) while the latter only applies to a part (17), and we adopted the results using the former method.
Reassuringly, the two methods yield consistent results for spectra that are applied to both.
Our main results are as follows.

1. The $\Delta\alpha/\alpha$ measured using the 86 spectra are all consistent with 0 by $<3\sigma$.
The ratio of $\Delta\alpha/\alpha$ to its error obeys the standard normal distribution, which supports the assumptions that the truth value is 0 and that our estimates of errors are reliable.

2. The weighted average value of $\Delta\alpha/\alpha$ agrees with 0, whether the whole sample or different subsamples is used.
If using all the 86 spectra, the average is $(-3\pm6)\times10^{-5}$.
The average values using the LAE and QSO subsamples have similar errors of $7\times10^{-5}$ and $1.1\times10^{-4}$, respectively.

3. The average $\Delta\alpha/\alpha$ at three redshifted ranges of $1.09<z<1.66$, $2.01<z<2.57$ and $3.21<z<3.73$ are $(5\pm3)\times10^{-4}$, $(-6\pm6)\times10^{-5}$ and $(4\pm5)\times10^{-4}$, respectively, all in accordance with 0.
We found no variation of $\alpha$ over time.
We limited the rate at which $\alpha$ varies to be $k=(-3\pm6)\times10^{-15}$ yr$^{-1}$ since $z=3.73$ using our measurements.
By combining our results with the measurements at $z<1$ using SDSS QSO spectra, we limited the rate to be $k=(0\pm4)\times10^{-15}$ yr$^{-1}$.

In addition, our results suggest that starburst galaxies' spectra have a better application prospect than QSO spectra in future studies of $\alpha$ variation.

\section*{Acknowledgements}

This work is based on observations obtained with the Very Large Telescope, programs 084.A-0303, 086.B-0320, 087.A-0610, 087.B-0229, 088.A-0672, 088.B-1034, 089.B-0275, 089.B-0936, 089.B-0951, 0909.A-0830, 090.B-0424, 091.A-0413, 091.B-0900, 092.A-0391, 092.B-0860, 093.A-0882, 093.B-0553, 094.B-0111, 095.B-0507, 096.A-0348, 097.A-0153, 098.B-0556, 099.A-0018, 099.A-0254, 099.A-0758, 099.B-0118, 101.A-0528, 101.B-0262, 101.B-0739, 101.B-0779, 102.A-0335, 102.A-0391, 102.A-0652, 103.A-0253, 103.A-0688, 103.B-0446, 104.A-0236, 189.A-0424.
Based on data products from observations made with ESO Telescopes at the Paranal Observatory under ESO programme ID 179.A-2005.

\section*{Data Availability}

The data underlying this article are available in the ESO data archive at http://archive.eso.org.

\begin{appendix}

\section{Details of Data Reduction}

\setcounter{figure}{0}
\renewcommand{\thefigure}{A\arabic{figure}}

We briefly introduce the pip-2D data generated by the XShooter pipeline.
The data is contained in fits files with three header data units (HDU).
As shown in Figure A1(a), the first HDU stores the 2D spectrum of a target (in the unit of ADU); the second stores the flux error of spectrum; the third stores the bad CCD map, marking those pixels affected seriously by bad CCD pixels, cosmic rays, and other factors.
The 2D spectrum had been straightened and aligned, with the wavelength direction going horizontally at a fixed interval of 0.6\AA  and the spatial direction going vertically.
This kind of 2D spectrum was produced by combining the 4 frames taken in 4 consecutive exposures under the ``ABBA'' dither mode.
Using an A$-$B$-$B$+$A algorithm, the pipeline eliminated the sky emission and left 3 images of the target on the combined frame, among which the bright image originated from the two exposures labelled as A and have positive fluxes and the other two dark images originated from the two exposures labelled as B and have negative fluxes.

We extracted 1-D spectrum based on the pip-2D data as follows.

First, the aperture for extracting the spectra from the three images was determined, as shown in Figure A1(b).
Almost all the targets are point-like sources, or close to point-like sources, and their brightness profiles along the spatial direction, cutting at the non-emission wavelengths for QSOs and at the \oiii $\lambda$5008 emission lines for LAEs, were well fitted by Gaussian functions.
The best-fitting Gaussian models give positions of the three images and the FWHMs of their brightness profiles, in which the latter were used to determine the width of the aperture for extracting the 1-D spectra.
The only exceptional target whose brightness distribution cannot be approximated by a Gaussian function is SA, and we will describe its extraction in another paragraph.

Second, spectra were extracted from the three images and merged into one spectrum, as shown in Figure A1(c).
Using the previously determined aperture width, we extracted the 1-D spectra from the three images, to which the first HDU contributed the flux and the second HDU the error.
With reference to the third HDU, we prepared a boolean value named ``mask'' for each wavelength of the 1-D spectra. The value was set to 0 if the data quality of all the pixels inside the aperture at this wavelength is 0, and to 1 otherwise.
The three spectra were then rescaled to eliminate the small differences between their fluxes caused by weather changes, and the negative fluxes were turned positive.
After the rescaling, new contaminated pixels were identified, at which the flux of one spectrum deviates from those of the other two spectra by more than 4.5$\sigma$.
These pixels were affected by cosmic rays or other adverse factors but had not been identified by the pipeline, so we changed their mask values to 1.
We calculated the weighted average of the three spectra using pixels with mask values of 0 and obtained a 1-D spectrum.
This averaged spectrum was also assigned a mask value at each wavelength: it was set to 1 if the mask values in the three spectra are all 1, and to 0 otherwise.
Besides, a few observations were not taken under the dither mode and each had only one image in its pip-2D data.
In these cases, the rescaling and the averaging were skipped.

Third, for each target, we calibrated the flux of its 1-D spectra and corrected the telluric absorption, as shown in Figure A1(d).
The flux of the 1-D spectra of a target was calibrated using the sensitivity function (the conversion from ADU to flux) stored in its flux-calibrated pip-1D data.
As telluric absorptions appeared in the targets' spectra, they must have also appeared in the spectra of standard stars observed close in time.
These standard stars' spectra, which belong to project 60.A-9022(C), can be obtained from the ESO archive.
So, to correct the telluric absorption in each target's spectrum, we built a telluric absorption model using the spectrum of the standard star, which was observed closest in time and of whose continuum the SNR is $>$300 in the wavelength range of the NIR arm.
For the spectrum of each standard star, we searched for a similar stellar spectrum in the XShooter Spectral Library \citep{Gonneau2020} and considered it as the template, and a telluric absorption model can be calculated by dividing the observed spectrum by the template.
Corrected QSO spectra using such telluric absorption models were examined in the continuum. The correction was successful at most of the wavelengths.
At some wavelengths with severe telluric absorption, the correction was less successful, which may be caused by the inhomogeneity or the variability in the properties of the atmosphere.

Finally, for each spectrum pixel, once it meets any of the following situations, we changed its mask value to 1.
The situations are:
1. The telluric absorption model of the pixel is below 0.5.
2. A strong SEL affects the pixel.
We adopted the list of SELs from \cite{Oliva2015}, and selected 184 strong lines with flux rates $>1000$.
With the information on these lines and with that revealed by the 1-D error spectrum of each target, we identified the sky emission lines in each spectrum and calculated their width for masking, in which an algorithm was involved to give greater widths for stronger lines.
3. The pixel lies near the \oiii doublet and is affected by visually identified cosmic rays that the pipeline had missed.
The final mask values label pixels affected by strong SELs, TALs, cosmic rays and CCD bad pixels.

We extracted a 1-D spectrum for each pip-2D data.
For a target, if there were more than one group of exposures taken under the dither mode and hence multiple pip-2D data in the ESO archive, we combined the extracted spectra, as we had done in combining spectra of the three images in the same pip-2D data.

\begin{figure}
\centering{
  \includegraphics[scale=0.8]{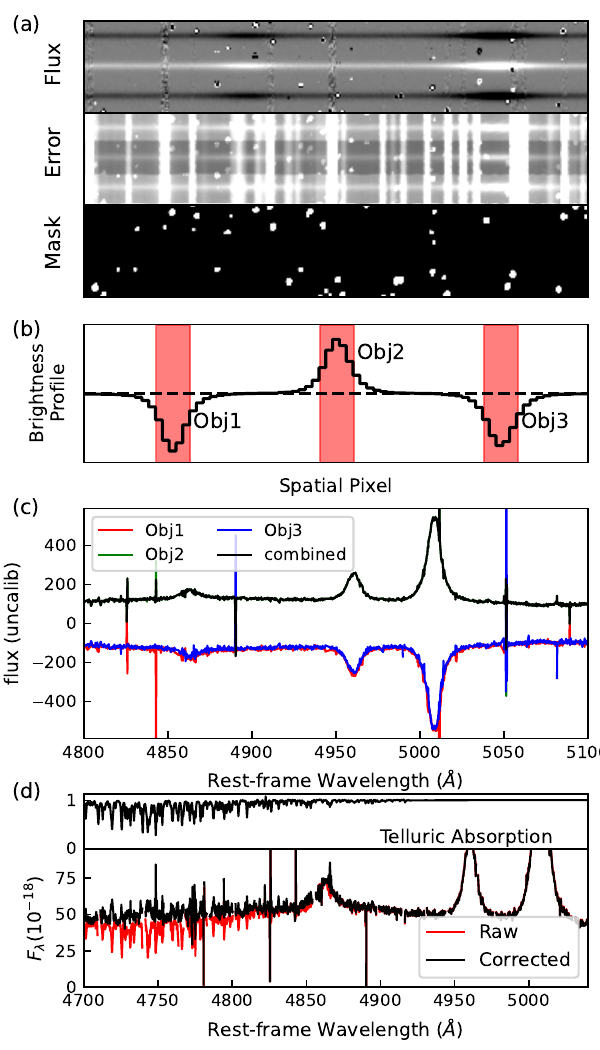}
  \caption{
  Example of a typical approach of extracting 1-D spectrum.
  {\bf (a)}: Display of pip-2D data.
  {\bf (b)}: Example of brightness distribution (black line) of QSO continuum and determination of apertures for extracting (red shade).
  {\bf (c)}: Example of combining the spectra of the three images.
  {\bf (d)}: Example of correction for telluric absorption.
  }}
  \label{figA1}
\end{figure}

SA is a lensed galaxy with multiple clumps, which could be star clusters.
The VLT/XShooter observation of SA has been described in detail by \cite{Vanzella2020}.
This observation contained 3 exposures, during which the slit was placed differently but covered at least two clumps each time.
The 2-D spectra of the three exposures are displayed in Figure A2.
We extracted 6 spectra using 6 apertures shown in the red line in the figure, denoted as SA1 to SA6.

\begin{figure}
\centering{
  \includegraphics[scale=0.8]{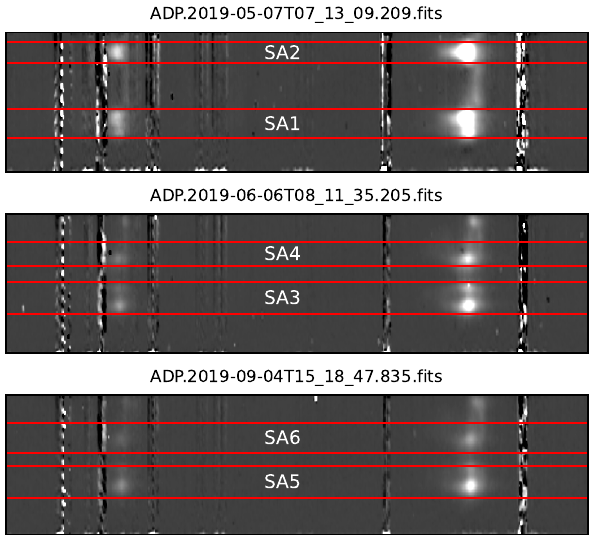}
  \caption{The apertures for extracting the spectra of SA.}}
  \label{figA2}
\end{figure}

\section{Applicability of Profile-Matching Method}

\setcounter{figure}{0}
\renewcommand{\thefigure}{B\arabic{figure}}

As we have demonstrated in section 3.1, the $\chi^2(\eta,A)$ surfaces obtained by the MG method are smooth, and the contours are close to ellipses, and the best estimates and statistical errors of $\eta$ and $A$ can be reliably obtained.
However, the results obtained by the PM method do not guarantee these properties.
We found two types of problems.
We illustrate each type with an example (Figure B1) and analyze the possible origin of the problem.

We show the first type of problem with LAE J0332-2746 as an example.
When the PM method was used, a feature similar to a geological fault occurs on the $\chi^2$ surface.
Accordingly, jumps can be seen on the $\chi^2(\eta)$ curve which are roughly evenly spaced with an interval of $\Delta\eta=\frac{0.6\AA }{\lambda_1}=\frac{1.2\times10^{-4}}{1+z}$.
Such jumps in the values of $\chi^2$ come from the changes in the pixels of the $\lambda$4960 line included in the matching.
When $\eta$ increases, the rightmost pixel is excluded, and at the opposite end, a new pixel joins in to become the new leftmost pixel.
The jumps are not obvious in the range of $\chi^2(\eta)<\chi^2_{\rm min}+1$ when the $\lambda$4960 line has high SNR and small FWHM, but they can be very prominent when the $\lambda$4960 line suffers from great noises, which will make the measurement of $\eta$ unreliable.

We then show the second type with LAE J1001+0206 as an example.
When measured by the PM method, wavy structures appear on the $\chi^2$ surface.
The origin of this structure is as follows.
Strong noise causes fault signals to be mistaken for structures on the line profiles.
When the fault signals from the two emission lines happen to be aligned, a local minimum appears on the $\chi^2(\eta)$ curve.
The width of the \oiii emission lines' profile plays an important role in the emergence of such wavy structures.
The broader the \oiii lines, the more likely the wavy structures will appear.

By visually examining the $\chi^2$ surfaces, we selected 17 spectra with which $\eta$ can be reliably measured by the PM method, including 16 LAE and 1 QSO spectra.
The \oiii emission lines in these spectra are narrow, have high SNR and are little affected by masks.
Notably, the FWHMs of these spectra's \oiii emission lines span within 16 pixels.
The previous work of \citep[e.g.,][]{Bahcall2004} used QSOs' SDSS spectra, in which the typical \oiii emission lines' FWMHs of 400-1000 km s$^{-1}$ span for 6--14 pixels.
These are informative to the applicability of the PM method.
We suggest that for spectra taken by Xshooter, SDSS or other spectrometers, as long as the \oiii emission lines' FWMHs span no more than 14--16 pixels, the application of the PM method can be considered.

\begin{figure}
\centering{
  \includegraphics[scale=0.8]{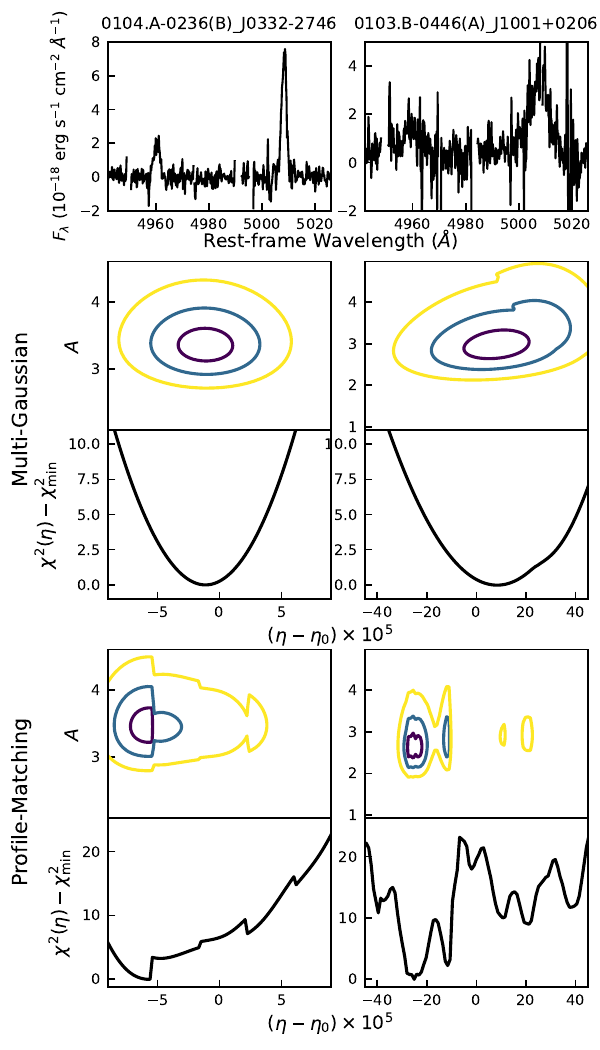}
  \caption{
  Two examples of spectra with which $\eta$ can be reliably measured by the MG method but not by the PM method.
  The top row shows the spectra, the middle two rows show the $\chi^2(\eta,A)$ surface and $\chi^2(\eta)$ curve obtained by the MG method, and the bottom two rows show those by the PM method.
  }}
  \label{figB1}
\end{figure}

\section{The Multi-Gaussian Fitting Results}

\setcounter{figure}{0}
\renewcommand{\thefigure}{C\arabic{figure}}

We present all 86 spectra and the best-fitting results using the multiple-Gaussian model in Figures C1, C2, C3 and C4.

\begin{figure*}
\centering{
  \includegraphics[scale=0.8]{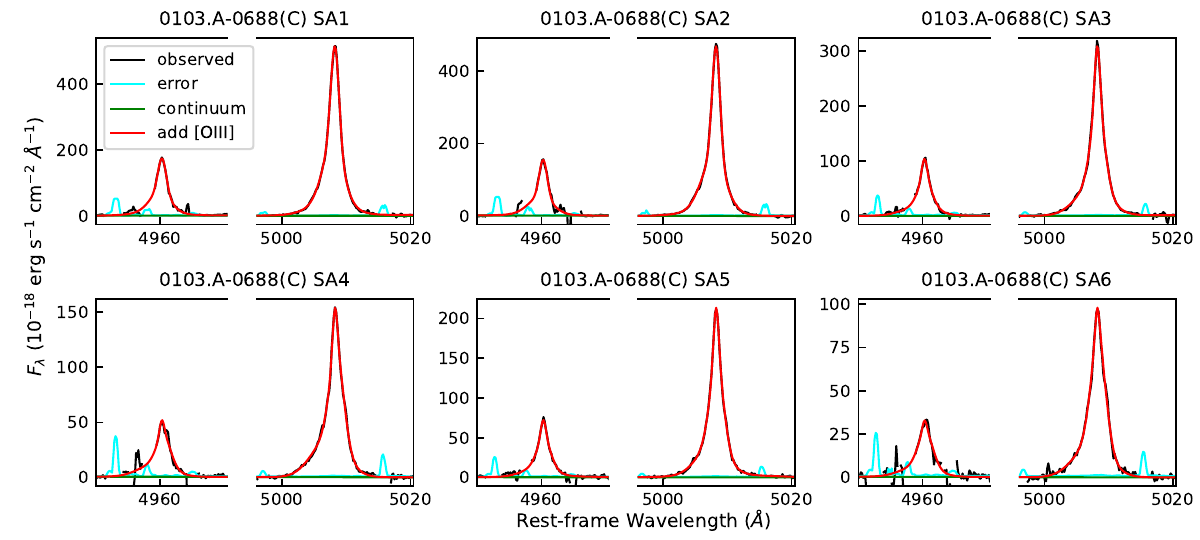}
  \caption{6 SA's spectra and best-fitting models.}}
  \label{figC1}
\end{figure*}

\begin{figure*}
\centering{
  \includegraphics[scale=0.85]{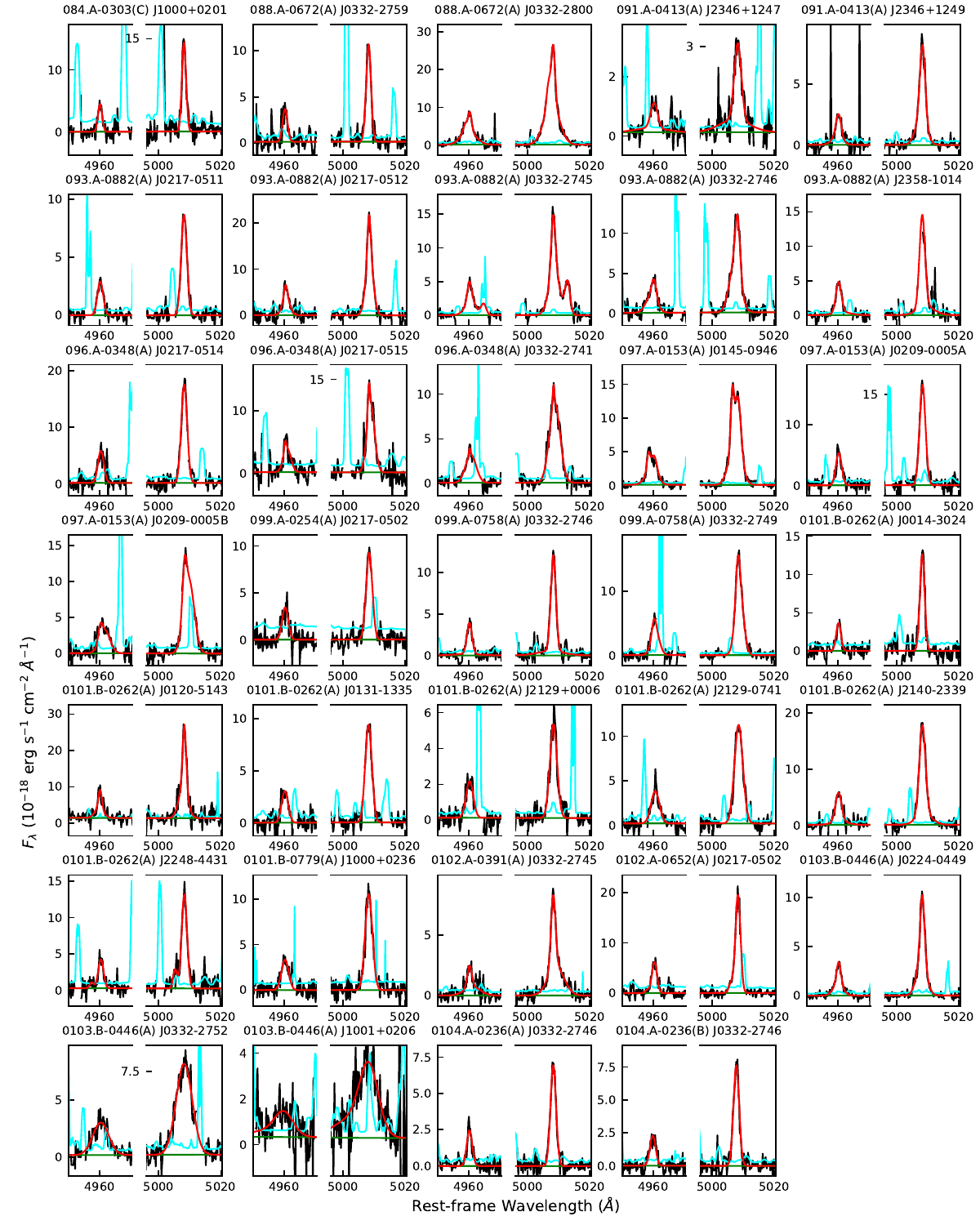}
  \caption{34 other LAEs' spectra and best-fitting models.}}
  \label{figC2}
\end{figure*}

\begin{figure*}
\centering{
  \includegraphics[scale=0.85]{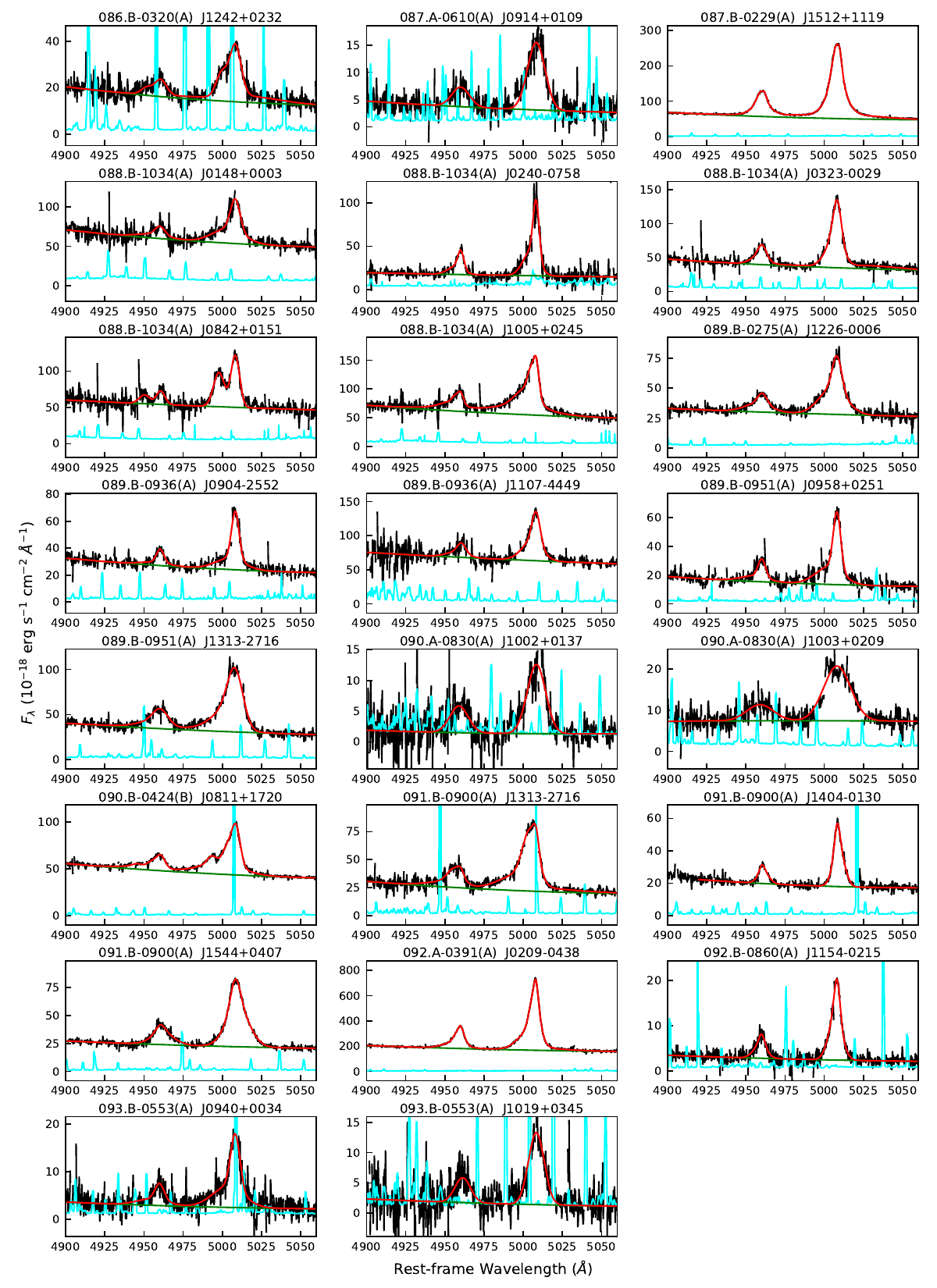}
  \caption{46 QSOs' spectra and best-fitting models (part 1).}}
  \label{figC3}
\end{figure*}

\begin{figure*}
\centering{
  \includegraphics[scale=0.85]{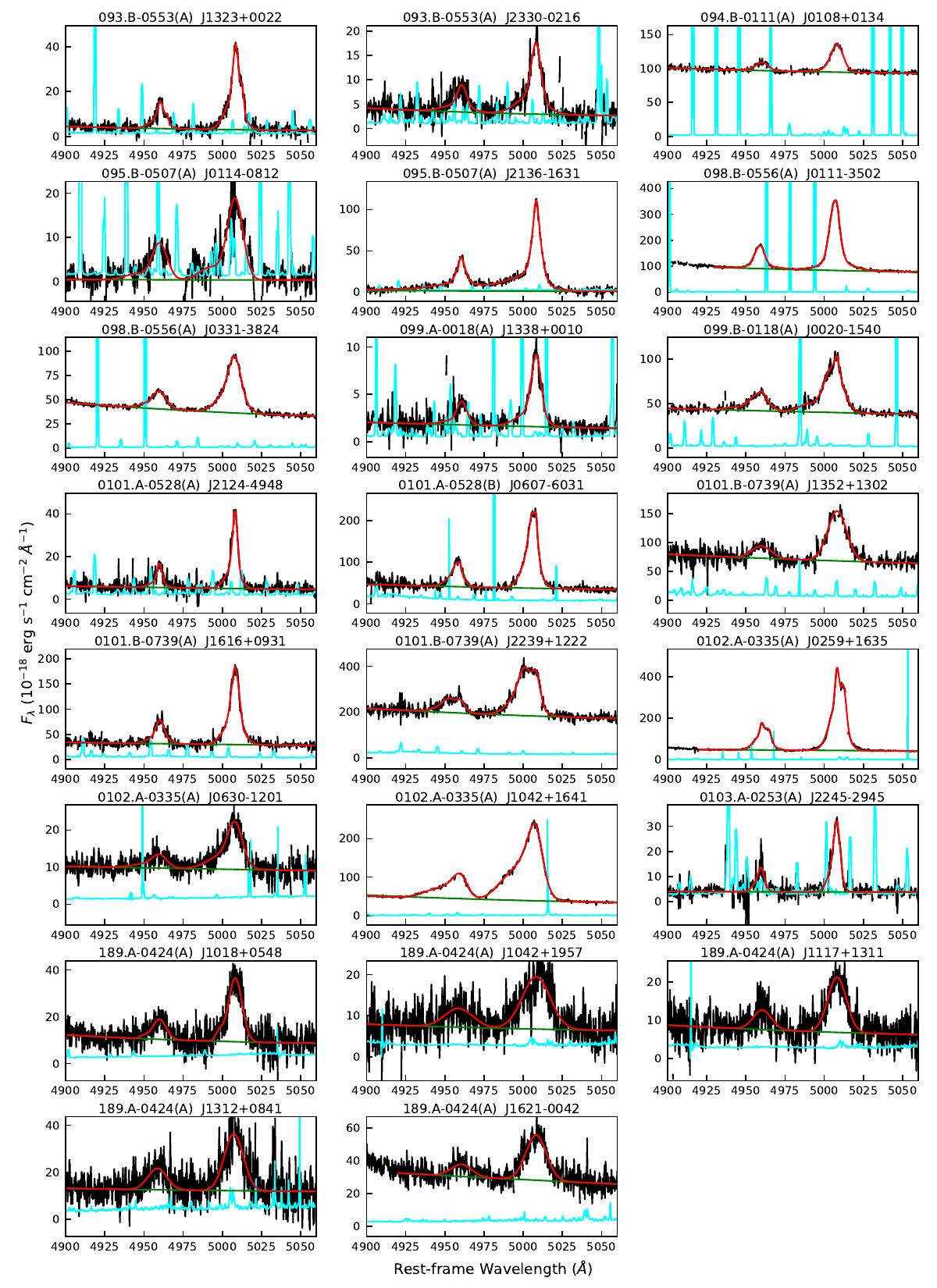}
  \caption{46 QSO spectra and best-fitting models (part 2).}}
  \label{figC4}
\end{figure*}

\section{Factors Affecting Measurement Precision of $\eta$}

\setcounter{figure}{0}
\renewcommand{\thefigure}{D\arabic{figure}}
\setcounter{equation}{0}
\renewcommand{\theequation}{D\arabic{equation}}

Factors that affect the precision of $\eta$ include the width and the SNRs of the \oiii doublet, and we explored their influence using simulated spectra.
Assuming that both \oiii lines have the same Gaussian profile, we generate spectra in which the FWHMs and SNRs of the \oiii doublet vary independently.
In a simulated spectrum, the FWMHs of the doublet were set to be equal, and the SNRs of the doublet, which were adjusted by assigning varied flux errors, are uncorrelated and can be different.
Measurements on $\eta$ using these simulated spectra revealed a relation between $\sigma_{\rm stat}(\eta)$ and the FWHM and SNRs of the doublet, which can be expressed as:
\begin{equation}
\sigma_{\rm stat}(\eta) \approx 1.9\times10^{-4} \left( \frac{\rm FWHM}{\rm 100\ km\ s^{-1}} \right)
  \sqrt{ \frac{1}{\rm SNR_{4960}^2} + \frac{1}{\rm SNR_{5008}^2} },
\end{equation}
This relation holds when the FWHM is 100--1200 km s$^{-1}$ and SNRs are 10--500.

We plot the observed \oiii SNRs of the final sample in Figure D1.
Figure D1 also shows the contour of the values of $\sigma_{\rm stat}(\eta)$ with the FWHM of the doublet fixed at 300 km s$^{-1}$.
In our sample, the values of ${\rm SNR}_{5008}$ are greater than those of ${\rm SNR}_{4960}$ for all the spectra, and their average ratio is 2.06.
By assuming that SNR$_{5008}=2.06 {\rm SNR}_{4960}$, we can simplify Equation D1 as:
\begin{equation}
\sigma_{\rm stat}(\eta) \approx 2.1\times10^{-6} \left( \frac{\rm SNR_{4960}}{100} \right)^{-1}
\left( \frac{\rm FWHM}{\rm 100\ km\ s^{-1}} \right).
\end{equation}
In the simulation, when we were adjusting the SNRs of the doublets by assigning varied flux errors, we found that the values of SNR follow such relation with the value of FWHMs and the flux errors: ${\rm SNR} \propto {\rm FWHM}^{-1/2} {\rm Err}$.
Hence if considering the influence of the FWHM alone, $\sigma_{\rm stat}(\eta)$ will be proportional to ${\rm FWHM}^{3/2}$.

\begin{figure}
\centering{
  \includegraphics[scale=0.8]{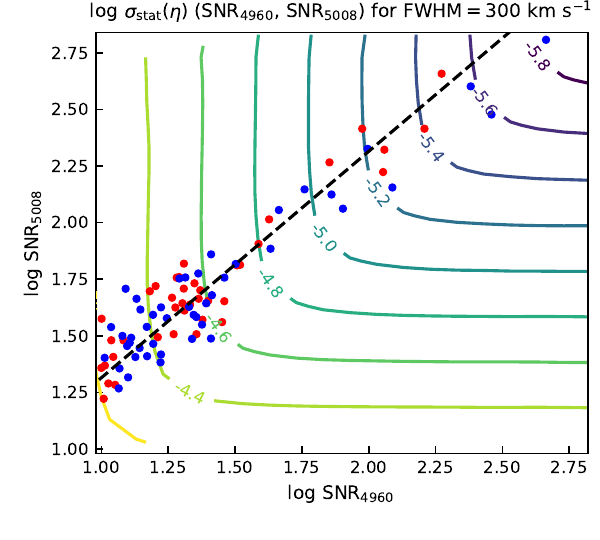}
  \caption{
  The contours show the simulated relation between $\sigma_{\rm stat}(\eta)$ and SNRs of the doublet lines when FWHM$=300$ km s$^{-1}$ is assumed.
  The data points show the observed \oiii SNRs in the final sample, red for LAEs and blue for QSOs.
  The dashed line shows the relation assuming SNR$_{5008}=2.06 {\rm SNR}_{4960}$.
  }}
  \label{figD1}
\end{figure}

The spectral resolution $R$ was generally considered to be one of the factors affecting the measurement precision.
Higher resolution leads to more accurate wavelength calibration because the calibrating emission lines of arc lamps and SELs will be narrower and more resolved, offering more information for finding the wavelength solution.
It then helps reduce the systematic error $\sigma_{\rm sys}(\eta)$ originating from the wavelength calibration.
Higher resolution also helps reduce the statistical error $\sigma_{\rm stat}(\eta)$, because the SELs and TALs will be narrower and affect fewer pixels, leading to an increase in the SNRs of \oiii lines and hence reducing $\sigma_{\rm stat}(\eta)$.
In addition, when the resolution is poor, the observed lines will appear broader.
Assuming that the intrinsic profile of a \oiii emission line is a Gaussian with a FWHM of ${\rm FWHM_{intr}}$, and assuming that the broadening function of spectrometry is also a Gaussian, the observed FWHM (${\rm FWHM_{obs}}$) of this line is roughly:
\begin{equation}
{\rm FWHM_{obs}^2 = FWHM_{intr}^2} + (c/R)^2,
\end{equation}
where c is the speed of light.
Higher resolution leads to smaller ${\rm FWHM_{obs}}$, and then smaller $\sigma_{\rm stat}(\eta)$.
Nevertheless, to what extent can $\sigma_{\rm stat}(\eta)$ be reduced by improving the resolution is limited by ${\rm FWHM_{intr}}$.
For Xshooter, the resolution $R$ is 4200--8100 (Tables 1 and 2), corresponding to $c/R$ of 37--70 km s$^{-1}$.
To a \oiii emission line with ${\rm FWHM_{intr}}$ of 100 –- 150 km s$^{-1}$,
the instrumental broadening from the Xshooter will bring an increase in ${\rm FWHM_{obs}}$ of about 10\%--30\%.
When the \oiii emission line's ${\rm FWHM_{intr}}$ is above 150 km s$^{-1}$, the increase will be no more than 10\%.
For our sample, the observed FWHMs of the LAE subsample are affected by $<$30\%, while those of the QSO subsample are almost unaffected.
In summary, increasing resolution benefits on improving the precision of $\eta$ measurement.
In future studies, if the LAE spectra are used for constraining $\alpha$ variation, we suggest that the resolution $R$ should be no less than 4000.
Otherwise, the observed width of the \oiii lines would be significantly larger than the intrinsic width, and the measurement precision would be unpromising.

\end{appendix}

\end{document}